\documentclass[12pt,titlepage]{article}
\usepackage{russlh}
\usepackage{amsthm,amsmath,amssymb,fullpage}
\usepackage[indentheadings]{russcorr}
\usepackage{russadd}

\theoremstyle{plain}
\newtheorem{theorem}{Теорема}[section]
\newtheorem{claim}[theorem]{Утверждение}
\newtheorem{lemma}[theorem]{Лемма}
\newtheorem{corollary}[theorem]{Следствие}
\theoremstyle{remark}
\newtheorem*{remark}{Замечание}
\theoremstyle{definition}
\newtheorem{algorithm}{Алгоритм}[section]
\newtheorem{definition}{Определение}[section]

\DeclareMathOperator{\dist}{dist} 
\DeclareMathOperator{\Probtmp}{Pr}
\newcommand{\Prob}[2][]{\Probtmp_{#1}\left[#2\right]}
\DeclareMathOperator{\col}{col} 
\DeclareMathOperator{\poly}{poly}
\DeclareMathOperator{\polylog}{polylog}
\DeclareMathOperator{\BPP}{BPP}

\newcommand{\N}{\mathbb{N}}
\newcommand{\B}{\mathcal{B}}
\newcommand{\bcube}[1]{\left\{0,\,1\right\}^{#1}}
\newcommand{\minen}{H_{\infty}}
\newcommand{\merg}[1]{\stackrel{#1}{\odot}}
\newcommand{\isub}[1]{\stackrel{#1}{\subset}}
\newcommand{\bydef}{:=}
\newcommand{\extr}[6]{($#2$,~$#3$)-экстрактор $#1\colon(#4)\times(#5)\to(#6)$}
\newcommand{\merger}[5]{$#2$-мёрджер $#1\colon(#3)^n\times(#4)\to(#5)$}

\begin{document}
\begin{titlepage}
\begin{center}
\large
Московский государственный университет им. М.\,В.\,Ломоносова\\
Механико-математический факультет\\
Кафедра математической логики и теории алгоритмов\\
\vspace{4cm}
Вступительный реферат в аспирантуру\\
\vspace{1.5cm}
\huge
Экстракторы и эффективный вариант\\
теоремы Мучника\\
\Large
\vspace{1cm}
Д.Мусатов\\
\large
\vspace{2cm}
Научный руководитель: д. ф.-м. н., профессор Н.\,К.\,Верещагин\\
\vspace{5.5cm}
Москва\\
2006
\end{center}
\end{titlepage}

\begin{abstract}
В 1999 году Ан.\,Мучник доказал следующую теорему: для любых двоичных слов $A$ и $B$ найдётся такое слово $X$ длины примерно $K(A|B)$, что
$K(X|A)\approx0$ и $K(A|B,\,X)\approx0$. $X$ выбирался как образ $A$ под действием хеш-функции, выбранной из некоторого небольшого
класса функций. Существование класса с нужными свойствами устанавливалось вероятностным методом. Ан.\,Мучник поставил вопрос: можно ли
построить такое семейство хеш-функций, чтобы хеш-значение вычислялось полиномиальным алгоритмом по слову $A$ и номеру функции? В работе
показано, что ответ на этот вопрос положительный, если немного ослабить формулировку исходной теоремы. А именно, в качестве семейства
хеш-функций надо взять экстрактор.

Экстрактором называется функция $Ext\colon\bcube{n}\times\bcube{d}\to\bcube{m}$, такая что для любого распределения вероятностей на
$\bcube{n}$ с большой минимальной энтропией и равномерного распределения на $\bcube{d}$ индуцированное распределение на $\bcube{m}$ близко к
равномерному. Понятие экстрактора появилось в начале 1990-х годов и нашло много приложений в различных областях теории сложности. В
литературе описаны различные конструкции экстракторов, обзор некоторых из них дан в настоящей работе. Явное построение экстрактора с
оптимальными параметрами, существование которого установлено вероятностным методом, является открытой проблемой.

Применение экстракторов не только отвечает на вопрос, поставленный Ан.\,Мучником, но и является инструментом для доказательства
аналогичных теорем для колмогоровской сложности с ограничением на ресурсы. В настоящей работе приведены первые результаты, полученные в этом
направлении: теорема для сложности с полиномиальным ограничением на память и теорема для сложности с полиномиальным ограничением на время,
где для декодирования используется алгоритм из класса AM.
\end{abstract}

\section{Введение}
В этом разделе дана мотивировка основных понятий теории экстракторов и кратко описаны полученные результаты об условном кодировании. Все
строгие определения и формулировки будут также даны в последующих разделах.
\subsection{Понятие экстрактора}
Решение многих задач в криптографии и других областях теории алгоритмов значительно упрощается за счёт использования в алгоритмах случайных
битов. При этом биты должны быть \лк действительно случайными\пк, то есть независимыми и принимающими значения $0$ и $1$ с вероятностями
$1/2$. Однако в природе такие идеальные распределения практически не встречаются. Например, если мы соорудим прибор, регистрирующий
случайные явления, возникающие на квантовом уровне, априори ниоткуда не следует независимость битов, которые он будет выдавать. Тем не
менее, хочется уметь применять вероятностные алгоритмы, имея в распоряжении только такие \лк квазислучайные\пк биты. Возникает вопрос: нельзя
ли алгоритмически \лк извлечь случайность\пк из \лк квазислучайного\пк распределения и получить таким образом некоторое меньшее число \лк
действительно случайных\пк или хотя бы \лк почти случайных\пк битов? Чтобы ответить на этот вопрос, нужно сначала формализовать понятие \лк
квазислучайности\пк. Наиболее естественным представляется следующее определение: будем говорить, что распределение \лк содержит $k$
случайных битов\пк, если его \textit{минимальная энтропия} не меньше $k$, \те вероятность любого элемента не превышает $2^{-k}$.

Пусть дано распределение на $\bcube{n}$ с минимальной энтропией $n-1$. Можем ли мы детерминированным алгоритмом извлечь хотя бы один \лк
действительно случайный\пк бит? Очевидно, нет! Ведь хотя бы для одного бита $b\in\bcube{}$ прообраз $b$ будет содержать больше $2^{n-1}$
слов, и если мы возьмём равномерное распределение на этом прообразе, то алгоритм заведомо выдаст $b$, и никакой случайности не будет.

Поэтому нам нужно ослабить требование. А именно, позволим нашему \лк извлекающему случайность\пк алгоритму использовать некоторое небольшое
число $d$ (\лк действительно\пк) случайных битов. Кроме того, ослабим требование к выходу алгоритма: будем требовать только, чтобы \лк
извлечённые\пк биты были \лк почти случайными\пк, то есть чтобы вероятность любого множества слов отличалась от его доли не больше, чем на
какое-то маленькое число $\eps$. Такие ослабленные требования приводят нас к классическому определению экстрактора:
\begin{definition}
Функция $Ext\colon\bcube{n}\times\bcube{d}\to\bcube{m}$ называется ($k$,~$\eps$)\д экстактором, если для любого распределения $X$ на
$\bcube{n}$ с минимальной энтропией не меньше $k$ и равномерного распределения $U$ на $\bcube{d}$ индуцированное распределение $Ext(X,\,U)$
$\eps$\д близко к равномерному на $\bcube{m}$.
\end{definition}
В некоторых приложениях не нужно, чтобы индуцированное распределение было близко к равномерному, а нужно лишь, чтобы почти все элементы в
образе имели ненулевую вероятность. В таком случае функцию называют \textit{дисперсером}. Часто экстракторы и дисперсеры рассматривают не
как функции, а как двудольные графы: левая доля отождествляется с $\bcube{n}$, правая\т с $\bcube{m}$, а множество рёбер, выходящих из
фиксированной вершины левой доли,\т с $\bcube{d}$. Более подробно понятие дисперсера и эквивалентность функций и графов рассмотрены в
разделе~\ref{extractors-pre}.

Однако встаёт вопрос: что мы можем получить от такого определения? Ведь нам всё равно нужны \лк действительно случайные\пк биты, пусть и в
меньшем количестве. Выясняется, что существуют экстракторы, для которых $d=O(\log n)$, и этого вполне достаточно для некоторых приложений.
Пусть, например, мы хотим симулировать некоторый алгоритм $A\in\BPP$, использующий $m$ случайных битов, при помощи $n$ \лк квазислучайных
битов\пк и экстрактора с подходящими параметрами. Можно показать, что с высокой вероятностью для большинства слов $y\in\bcube{d}$ алгоритм
$A$, получивший вместо случайных битов биты, выданные экстрактором, выдаст правильный ответ. А это значит, что мы можем \textit{перебрать}
все $2^d=\poly(n)$ вариантов слова $y$, для каждого из них запустить экстрактор, передать полученные биты алгоритму $A$ и выбрать наиболее
часто встретившийся ответ.

Чтобы использовать экстрактор для симуляции алгоритмов, работающих полиномиальное время, нужно чтобы сам экстрактор мог быть вычислен за
полиномиальное время. Вероятностным методом можно доказать существование для любых $k<n$ экстрактора с $d=\log n+O(1)$ и $m=k+d-O(1)$. В
статье~\cite{bounds} доказано, что это оптимальные параметры, то есть не может быть экстрактора с меньшим $d$ или большим $k$. Однако ни
одна из известных конструкций не достигает этих параметров. Все они имеют либо большее $d$, либо меньшее $m$, либо работают не для всех
значений $k$.

\subsection{Понятие колмогоровской сложности}
Пусть $A$ и $B$\т произвольные двоичные слова. Условной колмогоровской сложностью $K(A|B)$ слова $A$ отностительно слова $B$ называется
минимальная длина программы, переводящей $B$ в $A$. Колмогоровской сложностью $K(A)$ слова $A$ называется $K(A|\Lambda)$, где $\Lambda$\т пустое
слово, \те длина минимальной программы, порождающей слово $A$. При этом, разумеется, колмогоровская сложность зависит от способа задания
программ, однако различные естественные определения дают отличие сложности не более, чем на $O(\log n)$, для слов длины $n$. В формулировке
основной теоремы мы пренебрегаем таким различием, поэтому нам не нужно уточнять определение.

Также рассматривают колмогоровскую сложность с ограничением на ресурсы. Сложностью $C^{t,\,s}(A|B)$ называется минимальная длина программы,
переводящей $B$ в $A$ и работающей время $t$ на зоне $s$. В этом случае способ задания уже имеет значение, более того, имеет значение и
используемая модель вычислений. Так, различают сложности $C$ для детерминированных алгоритмов, $CN$ для недетерминированных, $CBP$ для
вероятностных и другие. Все необходимые точные определения даны в разделе~\ref{muchnik-pre}.

\subsection{Условное кодирование и теорема Мучника}
Рассмотрим такую задачу алгоритмической теории передачи информации. Пусть в точке $E$ известно некоторое слово $A$. Точку $E$ соединяет с
точкой $D$ канал ограниченной пропускной способности $k$, через который нужно передать некоторое сообщение $X$. В точке $D$ известно
некоторое другое слово $B$, и нужно по переданному сообщению $X$ расшифровать слово $A$. При этом операции зашифровки и расшифровки должны
выполняться алгоритмически. Вопрос: при каких условиях на $A$ и $B$ такое сообщение $X$ найлётся? Очевидное необходимое условие состоит в
том, что $k\geqslant K(A|B)$. Неожиданный результат, установленный Ан.\,Мучником в работе~\cite{muchnik} состоит в том, что это (с
определёнными оговорками) и достаточное условие. Это кажется удивительным: ведь на этапе кодирования мы ещё не знаем слова $B$, тем не менее
нам почти не нужно дополнительной информации для условного кодирования $A$. А именно, выполнена следующая 
\begin{theorem}[\cite{muchnik}]\label{muchnik-th}
Пусть $A$ и $B$\т двоичные слова длины не более $n$. Тогда найдётся такое слово $X$ длины не более $K(A|B)+O(\log n)$, что
$K(X|A)\leqslant O(\log n)$ и $K(A|B,\,X)\leqslant O(\log n)$.
\end{theorem}
Эту теорему можно сформулировать и так: среди всех программ, задающих $A$ при известном $B$, найдётся простая относительно $A$.
\begin{proof}[Идея доказательства]
Идея доказательства состоит в том, чтобы рассмотреть семейство хеш-функций, сопоставляющих словам длины $n$ слова длины $m=K(A|B)$, и взять
в качестве $X$ образ $A$ под действием одной из этих функций. Это
семейство должно обладать тем свойством, что для каждого слова $A$ найдётся хеш-функция, такая что по её значению $X$ и слову $B$ можно
легко (\те с логарифмическим числом дополнительной информации) восстановить $A$. Кроме того, это семейство должно быть полиномиального
размера, чтобы $X$ можно было задать номером соответствующей функции. Существование таких семейств хеш-функций устанавливается вероятностным
методом.
\end{proof}
Докладывая этот результат на Колмогоровском семинаре в Московском университете в 1999 году, Ан.\,Мучник поставил вопрос: возможно ли
построить \textit{полиномиально вычислимое} семейство хеш-функций с указанным свойством, \те можно ли вычислить хеш-значение по слову $A$ и
номеру функции за полиномиальное время? Настоящая работа частично отвечает на этот вопрос: да, можно, если в условии теоремы заменить
поправки $O(\log n)$ на $\polylog(n)$, при этом степень логарифма может быть сделана равной $2+\eps$ для любого $\eps>0$. А именно,
оказывается, что вместо хеш-функций, определённых Мучником, можно использовать экстракторы. Если будут построены оптимальные экстракторы,
вычислимые за полиномиальное время, то это сразу даст полный ответ на вопрос Мучника.

Применение экстракторов также позволяет доказать аналогичный результат для кодирования с несколькими условиями. Кроме того, данная техника
позволяет получать похожие теоремы для колмогоровской сложности с ограничением на ресурсы. В работе приведены первые результаты, полученные
в этом направлении:
\begin{theorem}
Найдётся такой полином $p_0=p_0(n)$, что для любого полинома $p\geqslant p_0$ найдутся полином $q$ и $t$, такие что
для любых слов $A$ и $B$ длины не более $n$, таких что $C^{\infty,\,p}(A|B)\leqslant k$, найдётся слово $X$ длины не более $k+O(1)$, такое что
$C^{t,\,q}(X|A)\leqslant O(\log^3 n)$ и $C^{\infty,\,q}(A|B,\,X)\leqslant O(\log^3 n)$.
\end{theorem}
\begin{theorem}
Для всякого полинома $p$ найдётся полином $q$, такой что для произвольных слов $A$ и $B$ длины не более $n$, таких что
$C^{p,\,\infty}(A|B)\leqslant k$, найдётся слово $X$ длины не более $k+O(\log^3 n)$, такое что $C^{q,\,\infty}(X|A)\leqslant O(\log^3 n)$ и
$CAM^{q,\,\infty}(A|B,\,X)\leqslant O(\log n)$.
\end{theorem}
Определения использованных сложностей даны в разделе~\ref{muchnik-pre}, доказательства теорем\т в
разделах~\ref{muchnik-polyzone-sect}~и~\ref{muchnik-polytime-sect}.

\subsection{Организация дальнейшего текста}
В разделе~\ref{preliminaries} даны определения экстрактора и доказаны простейшие факты. В разделе~\ref{implicit} вероятностным
методом доказано существование экстракторов с оптимальными параметрами. В разделах~\ref{explicit-first}--\ref{explicit-last} с разной
степенью подробности приведены некоторые конструкции полиномиально вычислимых экстракторов. В разделе~\ref{muchnik} при помощи экстракторов
доказаны теорема Мучника и её эффективные варианты.

\section{Основные определения и понятия теории экстракторов}\label{preliminaries}
\subsection{Минимальная энтропия и статистическое расстояние}
Для начала дадим два вспомогательных определения. Будем
рассматривать вероятностные распределения на конечных множествах.
Вероятность элемента~$a$ в распределении~$X$ будем обозначать
как~$X(a)$, вероятность множества~$S$\т как~$X(S)$.
\begin{definition}
Пусть дано вероятностное распределение~$X$ на конечном
множестве~$A$. Тогда его минимальной энтропией называется
величина $$H_{\infty}(X)=\min\limits_{a\in A}(-\log_2(X(a))).$$
\end{definition}
Таким образом, если $H_{\infty}(X)>k$, то все элементы множества~$A$
имеют вероятности меньше~$2^{-k}$.
\begin{definition}
Пусть даны вероятностные распределения~$X$ и~$Y$ на конечном
множестве~$A$. Тогда статистическим расстоянием между ними
называется величина
$$\dist(X,\,Y)=\frac12\|X-Y\|=\frac12\sum\limits_{a\in A}|X(a)-Y(a)|=\max\limits_{S\subset
A}|X(S)-Y(S)|.$$
\end{definition}
Будем говорить, что распределение~$X$ $\eps$-близко к~$Y$, если
$\dist(X,\,Y)\leqslant \eps$, иными словами, вероятности каждого
события по этим распределениям отличаются не более чем на~$\eps$.

\subsection{Экстракторы и дисперсеры}\label{extractors-pre}
Как уже отмечалось, экстракторы и дисперсеры можно рассматривать как функции и как двудольные графы. Приведём формальные определения.

Пусть $G$\т двудольный граф (возможно, с кратными ребрами) с
$N$~вершинами в левой доле, $M$~вершинами в правой доле и
степенью~$D$ каждой вершины из левой доли. Обозначим
через~$[N]=\{1,\dots,N\}$ множество вершин левой доли,
через~$[M]=\{1,\dots,M\}$\т правой, а через~$E$
множество ребер.

Пусть~$\Gamma(a)=\{z\in[M]\,|\,(a,\,z)\in E\}$\т множество всех
соседей вершины~$a$ из левой доли, $\Gamma(A)=\bigcup\limits_{a\in
A}\Gamma(a)$\т множество всех соседей подмножества~$A$ левой доли.

\begin{definition}
    Граф~$G=([N],\,[M],\,E)$ называется ($K$,~$\eps$)-дисперсером, если $\forall A\subset [N]$, $|A|\geqslant K$,
$|\Gamma(A)|\geqslant(1-\eps)M$.
\end{definition}

Это определение объясняет выбор названия \лк дисперсер\пк: все
достаточно большие подмножества левой доли имеют в соседях почти все
вершины правой доли, \те граф \лк рассеивает\пк (англ. disperse -
рассеивать) их по правой доле.

\begin{definition}
    Граф~$G=([N],\,[M],\,E)$ называется ($K$,~$\eps$)-экстрактором, если $\forall A\subset [N]$, $|A|\geqslant K$, $\forall B\subset [M]$
    $$\left|\frac{|E(A,\,B)|}{|A|\cdot D}-\frac{|B|}M\right|<\eps,$$
    где $E(A,\,B)$\т множество ребер, идущих из множества~$A$ в множество~$B$ в графе~$G$.
\end{definition}

Нетрудно заметить, что ($K$,~$\eps$)-экстрактор также является
($K$,~$\eps$)-дисперсером, достаточно взять $B=\Gamma(A)$.

\medskip Теперь дадим определение экстракторов и дисперсеров в терминах функций. Обозначим $\{0,\,1\}^l$ через $(l)$, а равномерное
распределение на этом множестве через $U_l$. Будем рассматривать функции $F:(n)\times(d)\rightarrow (m)$.

\begin{definition}
    Функция~$F:(n)\times(d)\rightarrow (m)$ называется ($k$,~$\eps$)-дисперсером, если для всех вероятностных   распределений~$X$ на~$\{0,\,1\}^n$ с~$H_{\infty}(X)\geqslant k$ и множеств $B\subset\{0,\,1\}^m$, $|B|\geqslant(1-\eps)M$
    $$\Prob{F(X,\,U_d)\in B}>0.$$
\end{definition}

\begin{definition}
    Функция~$F:(n)\times(d)\rightarrow (m)$ называется ($k$,~$\eps$)-экстрактором, если для всех вероятностных  распределений~$X$
на~$\{0,\,1\}^n$ с~$H_{\infty}(X)\geqslant k$
    $$\dist(F(X,\,U_d),\,U_m)<\eps.$$
\end{definition}

При таком определении ясен смысл названия \лк экстрактор\пк.
Функция~$G$ \лк извлекает\пк (англ. extract - извлекать) $m$~\лк почти
случайных\пк битов из поданных ей на вход  $n$~\лк квазислучайных\пк битов при
помощи $d$~\лк действительно случайных\пк битов.

Заметим, что если функция является экстрактором, то любой её префикс является экстрактором с теми же параметрами $k$ и $\eps$. Однако в
теореме Мучника полезно, чтобы префиксы являлись экстракторами с лучшими параметрами. Дадим следующее формальное определение.
\begin{definition}
    Обозначим через $X|_q$ префикс слова $X$ длины $q$. Функцию~$F:(n)\times(d)\rightarrow (m)$ назовём префиксным ($k$,~$\eps$)\д
экстрактором\footnote{Насколько известно автору, понятие префиксного экстрактора ранее не встречалось в литературе, однако соответствующее
свойство неоднократно отмечалось различными авторами.}, если для всех $i=0,\dots, k$ функция~$F|_{m-i}:(n)\times(d)\rightarrow (m-i)$,
определённая равенством
$F|_{m-i}(u,\,y)=(F(u,\,y))|_{m-i}$, является ($k-i$,~$\eps$)-экстрактором.
\end{definition}

Функции на словах и двудольные графы описанного вида естественным
образом соответствуют друг другу при $N=2^n$, $M=2^m$ и $D=2^d$. Левая доля графа отождествляется
со словами длины~$n$, правая\т со словами длины~$m$, а рёбра,
проведённые из фиксированной вершины левой доли,\т со словами
длины~$d$. При этом нумерация рёбер произвольна, \те разным функциям
может соответствовать один и тот же граф. Для префиксного экстрактора тоже можно и полезно рассмотреть соответствующий ему граф, но он не
будет иметь естественного определения в терминах графа, поскольку любое такое определение будет зависеть от нумерации вершин правой доли.
Кроме того, в отличие от обычного экстрактора определение префиксного не обобщается
на $N$, $M$, $D$ и $K$, не являющиеся степенями двойки.

\renewcommand{\labelenumi}{(\alph{enumi})}
\begin{claim}\label{equiv}
    Обозначим граф, построенный по функции~$F$, через $G_F$. Тогда
    \begin{enumerate}
        \item
        $F$ является ($k$,~$\eps$)-дисперсером $\Leftrightarrow$ $G_F$ является ($2^k$,~$\eps$)-дисперсером.
        \item
        $F$ является ($k$,~$\eps$)-экстрактором $\Leftrightarrow$ $G_F$ является ($2^k$,~$\eps$)-экстрактором.
    \end{enumerate}
\end{claim}
\begin{proof}
Переход от функций к графам очевиден: достаточно взять в качестве
распределения~$X$ равномерное на~$A$.

Обратный переход следует из того, что любое распределение с~$H_{\infty}(X)\geqslant k$
представляется в виде взвешенной суммы равномерных на множествах размера~$K$. Будем следовать доказательству этого факта, предложенному
М.\,Бабенко, а именно докажем по индукции следующее эквивалентное утверждение:
Пусть дано $N$ неотрицательных чисел, каждое из которых не превышает $1/K$ общей суммы. Назовём \лк операцией\пк одновременное уменьшение
$K$ чисел на одну и ту же величину (\лк величину операции\пк). Тогда не более чем за $N$ операций можно обратить все числа в $0$.

Выберем какие-нибудь $K$ ненулевых чисел. Уменьшим их на максимально возможную величину, так чтобы не нарушить условие. Тогда либо одно из
них обратится в нуль, и мы можем применить предположение индукции для $N-1$ и $K$ непосредственно, либо одно из невыбранных чисел ($x$)
станет равным $1/K$ общей суммы. В таком случае каждое из остальных чисел не превышает $1/(K-1)$ суммы остальных чисел, и мы можем для
остальных чисел применить предположение индукции для $N-1$ и $K-1$. А именно, добавим к каждому набору из предположения индукции число $x$ и
таким образом вместе с исходной операцией получим последовательность операций для исходного набора. Действительно, после первой операции
сумма всех чисел, кроме $x$, равна $(K-1)/K$ от суммы всех чисел. Значит, сумма всех величин операций из предположения индукции равна
$1/(K-1)\cdot(K-1)/K=1/K$ от общей суммы, то есть как раз $x$, что и нужно.
\end{proof}


\subsection{Полиномиально вычислимые экстракторы и постановка задачи оптимизации параметров}
Вероятностным методом можно доказать, что для всех $n$, $k$ и $\eps$ существуют экстракторы с $d=\log(n-k)+2\log(1/\eps)+O(1)$ и
$m=k+d-\log(1/\eps)-O(1)$. В работе~\cite{bounds} также показано, что это оптимальные параметры. Однако в большинстве приложений необходимо,
чтобы экстракторы были вычислимы за полиномиальное время. Определим формально, что это значит.
\begin{definition}
Пусть для функций $k(n)$, $\eps(n)$, $d(n)$, $m(n)$ задано семейство $Ext=\{Ext_n\}$ отображений
$Ext_n\colon\bcube{n}\times\bcube{d(n)}\to\bcube{m(n)}$. Тогда оно называется полиномиально вычислимым ($k$,~$\eps$)\д экстрактором, если
$Ext_n$ является ($k(n)$,~$\eps(n)$)\д экстрактором при всех $n$, и $Ext_n$ вычислима за полиномиальное от длины входа время.
\end{definition}
До сих пор неизвестно, можно ли построить полиномиально вычислимые экстракторы с оптимальными параметрами. Поэтому задача оптимизации
ставится так: либо при всех $k$ и $\eps$ достичь наибольшего возможного $m$ для оптимального $d$, либо достичь наименьшего возможного $d$
для оптимального $m$, либо достичь оптимальных $m$ и $d$ для некоторых значений $k$.

\section{Вероятностное доказательство существования экстракторов}\label{implicit}

Покажем, что случайный граф с определёнными параметрами с
положительной вероятностью является экстрактором.

\begin{theorem}[\cite{bounds}]\label{implicit-extr}
    Для всех $1<K\leqslant N$, $M>0$ и $\eps>0$ существуют:
    \begin{enumerate}
        \item ($K$,~$\eps$)-дисперсер для $D=\left\lceil\frac MK\left(\ln\frac1{\eps}+1\right)+\frac1{\eps}\left(\ln\frac
NK+1\right)\right\rceil$;
        \item ($K$,~$\eps$)-экстрактор для $D=\left\lceil\max\left\{\frac MK\cdot\frac{\ln2}{\eps^2},\ \frac1{\eps^2}\left(\ln\frac
NK+1\right)\right\}\right\rceil$;
	\item при условии, что $N=2^n$, $M$ и $K$ суть степени двойки, префиксный ($K$,~$\eps$)\д экстрактор для $\log
D=\left\lceil\log\left(\max\left\{\frac MK\cdot\frac{\ln2}{\eps^2},\ \frac1{\eps^2}\left(1+\ln2+\ln
N\right)\right\}\right)\right\rceil$.\footnote{Этого пункта теоремы не было в работе~\cite{bounds}.}
    \end{enumerate}
\end{theorem}
Прежде чем доказывать теорему, переформулируем её в терминах функций:
\begin{theorem}\label{implicit-extr-2}
Для всех $1\leqslant k \leqslant n$ и $\eps>0$ существуют:
\begin{enumerate}
\item
($k$,~$\eps$)\д дисперсер для $d=\log(n-k)+\log(1/\eps)+O(1)$ и $m=k+d-\log\log(1/\eps)-O(1)$;
\item
($k$,~$\eps$)\д экстрактор для $d=\log(n-k)+2\log(1/\eps)+O(1)$ и $m=k+d-2\log(1/\eps)-O(1)$;
\item
Префиксный ($k$,~$\eps$)\д экстрактор для $d=\log n+2\log(1/\eps)+O(1)$ и $m=k+d-2\log(1/\eps)-O(1)$.
\end{enumerate}
\end{theorem}
В работе~\cite{bounds} доказаны следующие нижние оценки:
\begin{theorem}[\cite{bounds}]
\begin{enumerate}
\item
Если функция $F\colon\bcube{n}\times\bcube{d}\to\bcube{m}$ является ($k$,~$\eps$)\д дисперсером, то $d \geqslant\log(n-k)+\log(1/\eps)-O(1)$
и $d+k-m \geqslant \log\log(1/\eps)-O(1)$.
\item\label{bounds-extr}
Если функция $F\colon\bcube{n}\times\bcube{d}\to\bcube{m}$ является ($k$,~$\eps$)\д экстрактором, то $d \geqslant\log(n-k)+2\log(1/\eps)-O(1)$
и $d+k-m \geqslant 2\log(1/\eps)-O(1)$.
\end{enumerate}
Таким образом, параметры, достигнутые в теореме~\ref{implicit-extr-2}, являются оптимальными.
\end{theorem}
Из оценки для экстракторов следует оценка для префиксных экстракторов: $d \geqslant\log n+2\log(1/\eps)-O(1)$. Таким образом, заявленные
параметры являются оптимальными.
\begin{proof}[Доказательство теоремы~\ref{implicit-extr}]
Докажем вначале оценку для дисперсеров. Возьмем двудольный
граф~$G=([N],\,[M],\,E)$ с $N$~вершинами в левой доле, $M$~вершинами
в правой, в котором из каждой вершины левой доли случайным образом
выпущено $D$~рёбер. Положим $L=\lceil\eps M\rceil$. Чтобы этот граф
не был ($K$,~$\eps$)-дисперсером, должны найтись подмножества
размера~$L$ в правой доле и размера~$K$ в левой, которые не
соединяет ни одно ребро. Таким образом, $\Prob{G \mbox{\ не\
является\ (}K,~\eps\mbox{)\д дисперсером}}$ не превосходит
\begin{equation}\label{implicit-dispersers}
C_N^K\cdot C_M^L\cdot\left(1-\frac
LM\right)^{KD}<\left(\frac{eN}K\right)^K\left(\frac{eM}L\right)^L\exp\left(-\frac{LKD}M\right).
\end{equation}
В этом переходе мы воспользовались неравенствами
$C_n^k<\left(e\frac nk\right)^k$, которое легко доказать по индукции, и
$1-x\leqslant e^{-x}$. Подставив $D$ из условия теоремы, получаем
\begin{multline}\exp\left(\frac{LKD}M\right)\geqslant\exp\left(L\left(\ln\frac1{\eps}+1\right)+K\cdot\frac LM\cdot\frac1{\eps}\left(\ln\frac NK+1\right)\right)\geqslant\\
\geqslant\exp\left(L\ln\frac {eM}L+K\ln\frac
{eN}K\right)=\left(\frac{eN}K\right)^K\left(\frac{eM}L\right)^L,
\end{multline}
откуда правая часть~\eqref{implicit-dispersers} не превосходит 1.
Значит, случайный граф с указанными параметрами не является
($K$,~$\eps$)-дисперсером с вероятностью меньше 1, \те является
($K$,~$\eps$)-дисперсером с положительной вероятностью, что и
требовалось доказать.

Теперь докажем оценку для экстракторов. Заметим вначале, что
изначальное требование, чтобы $\forall A\subset[N]$, $|A|\geqslant
K$ $\forall B\subset[M]$ выполнялось
$$\left|\frac{|E(A,\,B)|}{|A|\cdot D}-\frac{|B|}M\right|<\eps,$$
можно заменить на более слабое: $\forall A\subset[N]$, $|A|=K$
$\forall B\subset[M]$ $|E(A,\,B)|<KD\left(\frac{|B|}M+\eps\right)$.
Действительно, свед\'ение к множествам размера ровно~$K$ следует из
доказательства утверждения~\ref{equiv}. Далее, пусть найдутся
$A\subset[N]$ и $B\subset[M]$, такие что $|E(A,\,B)|\leqslant
KD\left(\frac{|B|}M-\eps\right)$. Тогда положим $\overline
B=[M]\setminus B$ и получим $|E(A,\,\overline B)|\geqslant
KD\left(\frac{|\overline B|}M+\eps\right)$, что противоречит
ослабленному требованию.

Далее, возьмём вновь двудольный граф~$G=([N],\,[M],\,E)$ с
$N$~вершинами в левой доле, $M$~вершинами в правой, в котором из
каждой вершины левой доли случайным образом выпущено $D$~рёбер.
Фиксируем $A\subset[N]$, $|A|=K$ и $B\subset[M]$. Положим
$p=\frac{|B|}M$. Оценим вероятность того, что $|E(A,\,B)|\geqslant
KD(p+\eps)$. Количество рёбер, идущих из $A$ в $B$, есть сумма
$KD$~независимых одинаково распределённых случайных величин,
принимающих значение 1 с вероятностью~$p$ и значение 0 с
вероятностью $1-p$. По неравенству Чернова мы можем оценить
$$\Prob{|E(A,\,B)|\geqslant KD(p+\eps)}\leqslant\exp(-2\eps^2KD).$$
Таким образом, $\Prob{G \mbox{\ не\ является\ (}K,~\eps\mbox{)\д
экстрактором}}$ не превосходит
\begin{multline*}
C_N^K\cdot2^M\exp(-2\eps^2KD)<\left(\frac{eN}K\right)^K 2^M\exp(-2\eps^2KD)=\\
=\left(e^{K(1+\ln(N/K))}\cdot
e^{-\eps^2KD}\right)\cdot\left(e^{M\ln2}\cdot e^{-\eps^2KD}\right).
\end{multline*}
Поскольку $D\geqslant\frac1{\eps^2}(1+\ln(N/K))$, первый множитель
не превосходит 1. Аналогично, поскольку
$D\geqslant\frac{M\ln2}{\eps^2K}$, второй множитель не превосходит
1. В итоге получаем, что
$$\Prob{G \mbox{\ не\ является\ (}K,~\eps\mbox{)\д экстрактором}}<1,$$ что и требовалось доказать.

Перейдём к префиксным экстракторам. Вначале переведём определение на язык графов. Назовём блоком уровня $i$ подмножество вершин правой доли
размера $2^i$, которое состоит из всех слов, имеющих фиксированный префикс длины $m-i$. Будем говорить, что множество $B$ $i$-вложено в
$[M]$ ($B\isub{i}[M]$), если любой блок уровня $i$ либо полностью входит в $B$, либо полностью не входит. Таким образом, $i$-вложенные
подмножества естественным образом соответствуют множествам слов длины $m-i$. Теперь можно определить префиксный экстрактор так: $\forall
i=0,\dots,k$ $\forall A\subset[N]$, $|A|=K/2^i$ $\forall B\isub{i}[M]$ $|E(A,\,B)|<\frac{KD}{2^i}\left(\frac{|B|}M+\eps\right)$.

Следуя схеме доказательства для экстракторов, получим, что
\begin{multline*}
\Prob{G \mbox{\ не\ является\ (}K,~\eps\mbox{)\д префиксным\ экстрактором}}\leqslant\\
\leqslant\sum\limits_{i=0}^k C_N^{K/2^i}\cdot2^{M/2^i}\exp(-2\eps^2KD/2^i)<\sum\limits_{i=0}^k \left(\frac{eN}{K/2^i}\right)^{K/2^i}
2^{M/2^i}\exp(-2\eps^2KD/2^i)=\\
=\sum\limits_{i=0}^k\left(e^{K/2^i(1+\ln(2^iN/K))}\cdot e^{-\eps^2KD/2^i}\right)\cdot\left(e^{M\ln2/2^i}\cdot
e^{-\eps^2KD/2^i}\right)\leqslant\\
\leqslant \sum\limits_{i=0}^k\cdot\left(e^{K/2^i(1+\ln N-\eps^2D)}\right)\cdot\left(e^{(M\ln2-\eps^2KD)/2^i}\right).
\end{multline*}
Второй множитель в каждом слагаемом, как и прежде, не превосходит 1. Поскольку $D\geqslant\frac1{\eps^2}(1+\ln2+\ln N)$, первый множитель в
$i$-ом слагаемом не превосходит $(1/2)^{K/2^i}$. Значит, сумма всех первых множителей меньше 1, значит и вся сумма меньше 1,
следовательно, исходная вероятность также меньше 1. Значит, граф с искомыми параметрами найдётся.
\end{proof}

\section{Экстракторы на базе хеш-функций}\label{explicit-first}
\subsection{Базовая конструкция}
\begin{definition}
Набор функций $H=\{h:[N]\rightarrow[L]\}$ называется семейством
хеш-функций со степенью коллизий $\delta$, если $\forall x_1\ne
x_2\in[N]$ $\Prob[h\in H]{h(x_1)=h(x_2)}\leqslant (1+\delta)/L$.
\end{definition}
Пусть функции $h \in H$ занумерованы произвольным образом. Тогда по
семейству~$H$ можно построить экстрактор по формуле $F(x,\,h)=h\cdot
h(x)$, где~$\cdot$~обозначает конкатенацию. Таким образом, $D=|H|$,
$M=DL$.
\begin{lemma}[\cite{ill}]
Пусть $H$\т семейство хеш-функций с вероятностью коллизий~$\delta$.
Тогда экстрактор, построенный по $H$, имеет параметры
$K=2^k=O(L/\delta)$ и $\eps=O(\sqrt{\delta})$
\end{lemma}
\begin{proof}
Определим вероятность коллизий для распределения~$X$ как
$\col(X)=\sum_a(X(a))^2$\т вероятность того, что независимо
выбранные в соответствии с~$X$ $x_1$ и $x_2$ совпадут. Покажем, что
$\col(X)\leqslant\frac1K$ при $H_{\infty}\geqslant k$.
Действительно,
$$\col(X)=\sum\limits_a(X(a))^2\leqslant2^{-k}\sum\limits_a X(a)=2^{-k}=\frac1K.$$ Далее, обозначим
через $Z$ распределение $h\cdot h(x)$, где $h$ распределено
равномерно, а $x$ в соответствии с $X$, и посчитаем $\col(Z)$.
$\col(Z)$\т вероятность того, что для независимо выбранных
$(h_1,\,x_1)$ и $(h_2,\,x_2)$ верно $h_1=h_2$ и $h_1(x_1)=h_2(x_2)$. Она
равна домноженной на $1/|H|$ вероятности того, что для
случайной~$h\in H$ и $x_1$, $x_2$, выбранных в соответствии с~$X$,
$h(x_1)=h(x_2)$. Вероятность того, что $x_1=x_2$, равна $\col(X)$.
Если же $x_1\ne x_2$, то эта вероятность не превосходит
$(1+\delta)/L$. В итоге получаем
\begin{equation}\label{colZ}
\col(Z)\leqslant\frac1{|H|}\left(\col(X)+\frac{1+\delta}L\right)\leqslant\frac1M+\frac1{|H|}\left(\frac1K+\frac{\delta}L\right).
\end{equation}
Далее,
$$\col(Z)=\sum\limits_a(Z(a))^2=\sum\limits_a\left(Z(a)-\frac1M\right)^2+\sum\limits_a\frac{2Z(a)}M-\sum\limits_a\frac1{M^2}=\sum\limits_a\left(Z(a)-\frac1M\right)^2+\frac1M,$$
откуда с учетом~\eqref{colZ} получаем
$\sum_a(Z(a)-1/M)^2\leqslant(1/K+\delta/L)/|H|$, откуда
$$\sum\limits_a\left|Z(a)-\frac1M\right|\leqslant\sqrt{M\left(\frac1K+\frac{\delta}L\right)\frac1{|H|}}=\sqrt{\frac LK+\delta},$$
что при $K=L/\delta$ влечёт $O(\sqrt{\delta})$-близость
распределения~$Z$ к равномерному, что и требовалось доказать.
\end{proof}

В~\cite{srinivasan-zuckerman} доказана следующая
\begin{lemma}
Для всех $1\leqslant L\leqslant N$ и $\eps>0$ можно построить
семейство $H$ хеш-функций, отображающих $[N]$ в $[L]$, с
вероятностью коллизий~$\eps$ размера $|H|=\poly(n,\,\eps^{-1},\,L)$.
\end{lemma}

Отсюда выводим
\begin{corollary}
Для всех $M\leqslant N$ и $\eps>0$ можно построить
($k$,~$\eps$)-экстрактор с~$D=\poly(n,\,\eps^{-1},\,M)$ и
$k=m-d+O(\log\eps^{-1})$.
\end{corollary}
Заметим, что построенный экстрактор требует, вообще говоря, очень
много случайных битов ($d$ полиномиально зависит от $m$), однако при
малых $m$ (точнее, при $m=\polylog(n)$) построенный экстрактор
является оптимальным.

\subsection{Композиция экстракторов}
\begin{definition}
Пусть $F_1:(n_1)\times(d_1)\rightarrow(m_1)$ и
$F_2:(n_2)\times(d_2)\rightarrow(d_1)$ суть экстракторы. Тогда
определим их композицию $F_1\circ
F_2:(n_1+n_2)\times(d_2)\rightarrow(m_1)$ по формуле $F_1\circ
F_2(x_1,\,x_2,\,y)=F_1(x_1,\,F_2(x_2,\,y))$.
\end{definition}
Таким образом, почти случайные биты, полученные с помощью второго
экстрактора, направляются на вход первому экстрактору в качестве
случайных. Разумеется, свойства первого экстрактора при этом могут
ухудшиться. Однако, если на вход композиции подать не произвольное
распределение~$X$, а распределение специального вида, то свойства
сохранятся.

Дадим формальное определение:
\renewcommand{\labelenumi}{\arabic{enumi}. }
\begin{definition}
Пусть $X_1$ и $X_2$ суть (определённые на одном вероятностном пространстве) случайные величины,
принимающие значения на~$\{0,\,1\}^{n_1}$ и~$\{0,\,1\}^{n_2}$
соответственно. Будем говорить, что они образуют
($k_1$,~$k_2$)-блочный источник, если
\begin{enumerate}
\item
$H_{\infty}(X_1)\geqslant k_1$.
\item
При любом фиксированном~$x_1$ $H_{\infty}(X_2|X_1=x_1)\geqslant
k_2$.
\end{enumerate}
\end{definition}
Это определение обобщается на произвольное число случайных величин.

Имеет место несложная
\begin{lemma}
Пусть $F_1:(n_1)\times(d_1)\rightarrow(m_1)$\т
($k_1$,~$\eps_1$)-экстрактор, а
$F_2:(n_2)\times(d_2)\rightarrow(d_1)$\т
($k_2$,~$\eps_2$)-экстрактор. Пусть также $(X_1,\,X_2)$\т
($k_1$,~$k_2$)-блочный источник. Тогда распределение $F_1\circ
F_2(X_1,\,X_2,\,U_{d_2})$ ($\eps_1+\eps_2$)-близко к равномерному.
\end{lemma}
\begin{proof}
Обозначим через $W$ случайную величину $F_2(X_2,\,U_{d_2})$.
Зафиксируем значение $x_1$, тогда при условии $X_1=x_1$
распределение $W$ $\eps_2$-близко к равномерному. Значит,
распределение пары $(X_1,\,W)$ $\eps_2$-близко к распределению
$(X_1,\,U_{d_1})$. Значит, распределение величины $F_1(X_1,\,W)$
$\eps_2$-близко к распределению $F_1(X_1,\,U_{d_1})$, которое, в
свою очередь, $\eps_1$-близко к равномерному, откуда распределение
$F_1(X_1,\,W)$ ($\eps_1+\eps_2$)-близко к равномерному, что и
требовалось.
\end{proof}

\subsection{Построение блочного источника}
Описаны различные методы построения блочных источников. Один из первых появился в работе~\cite{nisan-zuckerman} и основан на попарно
независимом выборе битов из исходного распределения. Другой метод описан в работе~\cite{nisan-ta-shma} и представлен в
разделе~\ref{ta-shma}. Этот метод был развит и усилен в работе~\cite{rsw}.

\section{Конструкция Та-Шмы}\label{ta-shma}

\subsection{Мёрджеры}

\begin{definition}
Случайная величина~$Z=Z_1\cdot\dots\cdot Z_b$ называется $b$-блочным
где-то случайным ($k$,~$\eps$,~$\eta$)-источником, если $Z_i$\т
случайные величины на~$\{0,\,1\}^k$ и существует случайная
величина~$Y$ на $\{0,\dots,b\}$, такая что:
\begin{itemize}
\item
Для всех $i\in\{1,\dots,b\}$ $\dist((Z_i|Y=i), U_k)\leqslant\eps$;
\item
$\Prob{Y=0}\leqslant\eta.$
\end{itemize}
$Y$ называется ($k$,~$\eps$,~$\eta$)-селектором для $Z$.
\end{definition}
Несложно доказать следующую лемму:
\begin{lemma}\label{srs}
\begin{enumerate}
\item

Любой где-то случайный ($k$,~$\eps$,~$\eta$)-источник
($\eps+\eta$)-близок к некоторому где-то случайному
($k$,~0,~0)-источнику.
\item
Если $Z$\т где-то случайный ($k$,~0,~0)-источник, то
$H_{\infty}(Z)\geqslant k$.
\end{enumerate}
\end{lemma}
Дадим определение мёрджера.
\begin{definition}
Функция $M\colon(k)^b\times(d)\rightarrow(m)$ называется
$\eps$-мёрджером, если для любого $b$-блочного где-то
случайного ($k$,~0,~0)-источника~$Z$ распределение $M(Z,\,U_d)$ $\eps$-близко к
равномерному.
\end{definition}

\subsection{Композиция двух экстракторов посредством мёрджера}
\begin{definition}
Пусть $E_1\colon(n)\times(d_1)\rightarrow(d_2)$ и
$E_2\colon(n)\times(d_2)\rightarrow(m_2)$\т экстракторы, а
$M\colon(m_2)^n\times(\mu_1)\rightarrow(m)$\т мёрджер. Тогда
композицией экстракторов посредством мёрджера называется функция
$E_2\stackrel{M}{\odot}E_1\colon(n)\times(d_1+\mu_1)\rightarrow(m)$,
определяемая следующим образом. Пусть $a\in\{0,\,1\}^n$,
$r_1\in\{0,\,1\}^{d_1}$, $r_2\in\{0,\,1\}^{\mu_1}$. Положим для
$i=1,\dots,n$ $q_i=E_1(a_{[i,\,n]}, r_1)$, а $z_i=E_2(a_{[1,\,i-1]},
q_i)$. Обозначим $E_2\ominus E_1=z_1\cdot\ldots\cdot z_n$ и положим
$E_2\stackrel{M}{\odot}E_1(a,\,r_1,\,r_2)=M(E_2\ominus E_1, r_2)$.
\end{definition}
\begin{remark}
Вообще говоря, слова $a_{[i,\,n]}$ и $a_{[1,\,i-1]}$ короче $n$
битов. Однако, поскольку мы рассматриваем распределения на этих
словах, и интересуемся только их минимальной энтропией, мы можем
формально дополнить эти слова до нужной длины, например нулями.
\end{remark}
\begin{remark}
Можно считать, что $m_1\geqslant d_2$, поскольку свойство близости к
равномерному распределению сохраняется при взятии ограничения на
часть битов.
\end{remark}
\begin{theorem}\label{twoextcomp}
Пусть $E_1$\т ($k_1$, $\zeta_1$)-экстрактор, $E_2$\т ($k_2$,
$\zeta_2$)-экстрактор, а $M$\т $\zeta_3$-мёрджер. Тогда для любого
параметра $s>0$ $E_2\merg{M} E_1$\т ($k_1+k_2+s$,
$\zeta_1+\zeta_2+\zeta_3+8n2^{-s/3}$)-экстрактор.
\end{theorem}
\begin{proof}
Очевидно, достаточно доказать, что $E_2\ominus E_1$\т где-то случайный ($m_2$,
$\zeta_1+\zeta_2$, $8n2^{-s/3}$)-источник. Пусть
$X$\т случайная величина с $H_{\infty}(X)\geqslant k_1+k_2+s$.
Обозначим через $Q_i$ и $Z_i$ случайные величины, принимающие
значения $q_i$ и $z_i$ соответственно. Положим $\eps_3=2^{-s/3}$,
$\eps_2=2\eps_3$, $\eps_1=2\eps_2$.

Определим селектор для величины $Z=Z_1\cdot Z_2\cdot\dots\cdot
Z_n=E_2\ominus E_1$. Пусть $w\in\{0,\,1\}^n$, и
$$f(w)=\max\{i\colon \Prob{X_{[i,\,n]}=w_{[i,\,n]}|X_{[1,\,i-1]}=w_{[1,\,i-1]}}\leqslant(\eps_2-\eps_3)\cdot2^{-k_1}\}$$
Это уже почти селектор, но его надо немного подправить, избавившись
от слишком редко принимаемых значений. Ведь если значение
принимается редко, то соответствующее условное распределение может
вести себя как угодно, а не быть близким к равномерному. Более
строго: назовём $w$ плохим, если $f(w)=i$ и
\begin{enumerate}
\item
$\Prob{f(x)=i}\leqslant\eps_1$, или
\item
$\Prob{f(x)=i\mid x_{[1,\,i-1]}=w_{[1,\,i-1]}}\leqslant\eps_2$, или
\item
$\Prob{x_i=w_i\mid x_{[1,\,i-1]}=w_{[1,\,i-1]}}\leqslant\eps_3$.
\end{enumerate}
Обозначим через $B$ множество всех плохих $w$, а через $B_i$\т
множество всех $w$, удовлетворяющих условию $i$. Теперь определим
селектор как
$$Y(w)=\begin{cases}
0,&\text{если $w$ плохое;}\\
f(w),&\text{иначе.}
\end{cases}$$
Нетрудно доказать, что доля плохих $w$ не превосходит
$n(\eps_1+\eps_2+\eps_3)\leqslant 8n2^{-s/3}$. Осталось доказать,
что $(Z_i|Y=i)$ ($\zeta_1+\zeta_2$)-близко к равномерному. Это
следует из двух утверждений:
\begin{claim}\label{claim41}
Если $\Prob{Y=i\mid X_{[1,\,i-1]}=w_{[1,\,i-1]}}>0$, то
$\minen(X_{[i\,n]}\mid Y=i \text{ и }
X_{[1,\,i-1]}=w_{[1,\,i-1]})\geqslant k_1$.
\end{claim}
\begin{claim}\label{claim42}
$\minen(X_{[1,\,i-1]}\mid Y=i)\geqslant k_2$.
\end{claim}
Действительно, для всех $w_{[1,\,i-1]}$, удовлетворяющих условию
утвеждения~\ref{claim41}, распределение $Q_i|Y=i \text{ и }
X_{[1,\,i-1]}=w_{[1,\,i-1]}$ $\zeta_1$-близко к равномерному (по
свойству экстрактора $E_1$). Отсюда распределение
$(X_{[1,\,i-1]}|Y=i)\times(Q_i|Y=i \text{ и }
X_{[1,\,i-1]}=w_{[1,\,i-1]})$ $\zeta_1$-близко к
$(X_{[1,\,i-1]}|Y=i)\times U_{d_2}$. Отсюда по свойству экстрактора
$E_2$ получаем, что $(Z_i|Y=i)$ ($\zeta_1+\zeta_2$)-близко к
равномерному.
\end{proof}
Докажем теперь утверждения~\ref{claim41} и~\ref{claim42}
\begin{proof}[Доказательство утверждения~\ref{claim41}]
Для любого $w$, такого что $Y(w)=i$, выполнено
\begin{multline*}
\Prob{x_{[i\,n]}=w_{[i\,n]}\mid x_{[1,\,i-1]}=w_{[1,\,i-1]},\ Y(x)=i}
\leqslant\frac{\Prob{X_{[i\,n]}=w_{[i\,n]}\mid x_{[1,\,i-1]}=w_{[1,\,i-1]}}}{\Prob{Y(x)=i\mid X_{[1,\,i-1]}=w_{[1,\,i-1]}}}\leqslant\\
\leqslant\frac{(\eps_2-\eps_3)\cdot2^{-k_1}}{\Prob{Y(x)=i\mid x_{[1,\,i-1]}=w_{[1,\,i-1]}}}
\leqslant\frac{(\eps_2-\eps_3)\cdot2^{-k_1}}{\eps_2-\eps_3}=2^{-k_1}
\end{multline*}
Первое неравенство следует из того, что
$\Prob{A|B}\leqslant\Prob{A}/\Prob{B}$, второе\т из того, что
$f(w)=i$, и определения $f$. Докажем третье, \те что
$\Prob{Y(x)=i\mid x_{[1,\,i-1]}=w_{[1,\,i-1]}}\geqslant\eps_2-\eps_3$,
если только не равно нулю. Действительно, если $w_{[1,\,i-1]}$
служит началом для некоторого $w$ с $Y(w)=i\ne0$, то никакое
продолжение $w_{[1,\,i-1]}$ не может быть плохим по первому и
второму условию. Значит,
$\Prob{Y(x)=i\mid x_{[1,\,i-1]}=w_{[1,\,i-1]}}=\Prob{f(x)=i\mid x_{[1,\,i-1]}=w_{[1,\,i-1]}}-\Prob{f(x)=i
\text{ и } x\in
B_3\mid x_{[1,\,i-1]}=w_{[1,\,i-1]}}\geqslant\eps_2-\eps_3$. Первое
слагаемое не меньше $\eps_2$, поскольку $x\not\in B_2$. Оценка на
второе тоже понятна: для тех $x$, где $f(x)=i$, соответствующая
вероятность не превосходит $\eps_3$ по определению $B_3$, а для
остальных и вовсе равна нулю.
\end{proof}
\begin{proof}[Доказательство утверждения~\ref{claim42}]
Пусть $w_{[1,\,i-1]}$\т произвольное слово, продолжаемое до $w$ с
$Y(w)=i$. Оценим вероятность $\Prob{x_{[1,\,i-1]}=w_{[1,\,i-1]}}$.
\begin{multline*}
\Prob{x_{[1,\,i-1]}=w_{[1,\,i-1]}}=\frac{\Prob{x_{[1,\,n]}=w_{[1,\,n]}}}{\Prob{x_{[i,\,n]}=w_{[i,\,n]}\mid x_{[1,\,i-1]}=w_{[1,\,i-1]}}}=\\
\frac{\Prob{x_{[1,\,n]}=w_{[1,\,n]}}}{\Prob{x_i=w_i\mid x_{[1,\,i-1]}=w_{[1,\,i-1]}}\cdot\Prob{x_{[i+1,\,n]}=w_{[i+1,\,n]}\mid
x_{[1,\,i]}=w_{[1,\,i]}}}
\end{multline*}
Числитель оценивается сверху как $2^{-(k_1+k_2+s)}$, поскольку
$\minen(X)\geqslant k_1+k_2+s$. Первый сомножитель знаменателя
оценивается снизу как $\eps_3$, поскольку $w\not\in B_3$. Наконец,
второй сомножитель знаменателя оценивается снизу как
$(\eps_2-\eps_3)\cdot2^{-k_1}$, поскольку $f(w)=i$. В итоге имеем
\begin{equation}\label{boundforprob}
\Prob{x_{[1,\,i-1]}=w_{[1,\,i-1]}}\leqslant\frac{2^{-k_2-s}}{\eps_3(\eps_2-\eps_3)}.
\end{equation}
Далее,
\begin{multline*}
\Prob{x_{[1,\,i-1]}=w_{[1,\,i-1]}\mid Y(x)=i}\leqslant\frac{\Prob{x_{[1,\,i-1]}=w_{[1,\,i-1]}}}{\Prob{Y(x)=i}}\leqslant\\
\leqslant\frac{2^{-k_1-s}}{\eps_3(\eps_2-\eps_3)\Prob{Y(x)=i}}\leqslant
\frac{2^{-k_1-s}}{\eps_3(\eps_2-\eps_3)(\eps_1-\eps_2-\eps_3)}=2^{-k_1}
\end{multline*}
Последнее неравенство следует из того, что если $\Prob{Y(x)=i}>0$,
то $\Prob{f(x)=i}\geqslant\eps_1$. Исключив $x\in B_2$ и $x\in B_3$,
получаем нужную оценку $\Prob{Y(x)=i}\geqslant
\eps_1-\eps_2-\eps_3$.
\end{proof}

\subsection{Композиция нескольких экстракторов}
Распространим нашу технику на произвольное число экстракторов.
\begin{definition}
Пусть $E_i\colon(n)\times(d_i)\to(d_{i+1}+s_{i+1})$ суть ($k_i$,
$\zeta_i$)-экстракторы для $i=1,\dots,t$, $s_i\geqslant0$, а
$s_2=0$. Пусть также
$M_i\colon(d_{i+2}+s_{i+2})^n\times(\mu)\to(d_{i+2})$ суть
$\bar\zeta_i$-мёрджеры для $i=1,\dots,t-1$. Определим функцию
$E=E_t\merg{M_{t-1}}E_{t-1}\merg{M_{t-2}}\cdots\merg{M_1}E_1\colon(n)\times(d_1+\mu_1+\dots+\mu_{t-1})\to(d_{t+1})$
индуктивно по правой ассоциативности: $E\bydef
E_t\merg{M_{t-1}}\left(E_{t-1}\merg{M_{t-2}}\cdots\merg{M_1}E_1\right)$.
\end{definition}
\begin{theorem}\label{manyextcomp}
Для любого параметра безопасности $s>0$ $E$ является ($\sum_{i=1}^t
k_i+(t-1)s$,
$\sum_{i=1}^t\zeta_i+\sum_{i=1}^{t-1}\bar\zeta_i+(t-1)n2^{-s/3+3}$)-экстрактором.
Если $E_i$ и $M_i$ вычислимы за полиномиальное время, то и $E$ тоже.
\end{theorem}
\begin{proof}
Параметры экстрактора очевидным образом получаются применением по
индукции теоремы~\ref{twoextcomp}. Докажем сохранение полиномиальной
вычислимости. Будем действовать при помощи динамического
программирования по следующему алгоритму:
\begin{enumerate}
\item
Вход: $x\in\bcube{n}$, $y\in\bcube{d_1}$ и $y_j\in\bcube{\mu_j}$,
$j=1,\dots,t-1$.
\item
Будем вычислять матрицу $M$ с элементами
$M_{ji}=\left(E_j\merg{M_{j-1}}\cdots\merg{M_1}E_1\right)\left(x_{[i,\,n]},
yy_1\dots y_{j-1}\right)$ для $1\leqslant i\leqslant n$ и
$1\leqslant j\leqslant t$.

Первый ряд матрицы, $M_{1i}$, может быть вычислен непосредственно
как $E_1(x_{[i,\,n]}, y)$. Пусть мы заполнили $j$-ый ряд матрицы,
заполним ($j+1$)-ый.
\begin{itemize}
\item
Обозначим $q_l=M_{jl}$, $l=i,\dots,n$, и положим
$z_l=E_{j+1}(x_{[i,\,l-1]},\,q_l)$.
\item
Положим $M_{j+1,l}=M_j(z_i\dots z_n,\,y_j)$.
\end{itemize}
\end{enumerate}
Результат вычислений будет правильным по определению мёрджера,
полиномиальность времени работы понятна.
\end{proof}

\subsection{Построение мёрджеров}
Опишем конструкцию явного построения мёрджеров. Вначале заметим, что
любой ($k$,~$\eps$)-экстрактор с $n=2k$ \лк извлекает случайность\пк
из любой случайной величины $X$ с $\minen(X)\geqslant k$, в
частности, из двухблочного где-то случайного ($k$,~0,~0)-источника.
(по лемме~\ref{srs}) Таким образом, он является двухблочным
мёрджером. Построим $b$-блочный мёрджер на базе двухблочных. А
именно, будем действовать так:
\begin{algorithm}\label{merger-constr}
Пусть $M\colon(k)^2\times(d(k))\to(k-m(k))$\т мёрджер. Построим
рекурсивно мёрджер $M_l\colon(k)^{2^l}\times(l\cdot
d(k))\to(k-l\cdot m(k))$:
\begin{enumerate}
\item
Вход: $x^l=x_1^l\dots x_{2^l}^l$, где $x_i^l\in\bcube{k}$;
$d=d_1\dots d_l$, $d_i\in\bcube{d(k)}$.
\item
Если $l=0$, возвращаем $x^l$.
\item
Иначе положим $x^{l-1}_i=M(x^l_{2i-1},\,x^l_{2i},\,d_l)$,
$i=1,\dots,2^{l-1}$.
\item
Возвратим $M_{l-1}(x^{l-1}_1\dots x^{l-1}_{2^{l-1}},\,d_1\dots
d_{l-1})$.
\end{enumerate}
\end{algorithm}
Докажем корректность работы алгоритма.
\begin{theorem}\label{mergexist}
Пусть для всех $k$ для некоторых монотонно растущих функций $d$, $m$
и $\eps^{-1}$ существует полиномиально вычислимый $\eps(k)$-мёрджер
$M\colon(k)^2\times(d(k)\to(k-m(k))$. Тогда $M_l$, построенный по
алгоритму~\ref{merger-constr}, является
$l\cdot\eps(k-m(k))$-мёрджером.
\end{theorem}
\begin{proof}
Для $j=l,\dots,0$ и $i=1,\dots,2^j$ обозначим через $X^j_i$
случайную величину, принимающую значения $x^j_i$ с вероятностями,
индуцированными распределением $X$ на $x=x^l\in\bcube{k\cdot2^l}$ и
равномерным на $d\in\bcube{l\cdot d(k)}$. Заметим, что $X^l=X$\т
исходное распределение, а $X^0$\т выход. Через $X^j$ обозначим
конкатенацию $X^j=X^j_1\dots X^j_{2^j}$ Обозначим $k_j=k-(l-j)m(k)$
и докажем более общий факт: если $X$\т где-то случайный
($k$,~0,~0)-источник, то для всех $1\leqslant i\leqslant2^j$
$$\dist((X^j_i|Y\in[2^{l-j}(i-1)+1,\,2^{l-j}_i]),\,U_{k_j})\leqslant(l-j)\eps(k_j),$$
где $Y$\т ($k$,~0,~0)-селектор для $X$. Проведём доказательство
нисходящей индукцией по~$j$. При $j=l$ утверждается, что $\forall
i\dist((X_i|Y=i),\,U_k)=0$, что верно по определению $Y$. Пусть мы
доказали утверждение для $j$, докажем для $j-1$. По предположению
индукции выполнено:
\begin{itemize}
\item
$\dist((X^j_{2i-1}|Y\in[2^{l-j}(2i-2)+1,\,2^{l-j}_{2i-1}]),\,U_{k_j})\leqslant(l-j)\eps(k_j)$;
\item
$\dist((X^j_{2i}|Y\in[2^{l-j}(2i-1)+1,\,2^{l-j}_{2i}]),\,U_{k_j})\leqslant(l-j)\eps(k_j)$.
\end{itemize}
Воспользуемся следующей леммой:
\begin{lemma}\label{concat}
Пусть $A$, $B$ и $Y$ суть случайные величины. Предположим, что
$\dist((A|Y\in S_1),\,U_k)\leqslant\eps$ и $\dist((B|Y\in
S_2),\,U_k)\leqslant\eps$ для некоторых непересекающихся множеств
$S_1$ и $S_2$. Тогда распределение $(AB|Y\in S_1\cup S_2)$
$\eps$-близко к некоторому распределению $W$ с $\minen(W)\geqslant
k$.
\end{lemma}
По лемме имеем, что
$(X^j_{2i-1}X^j_{2i}|Y\in[2^{l-j}(2i-2)+1,\,2^{l-j}_{2i}])$
$(l-j)\eps(k_j)$-близко к некоторому $W$ с $\minen(W)\geqslant k_j$.
Поскольку $X^{j-1}_i=M(X^j_{2i-1}X^j_{2i},\,d_j)$,
$(X^{j-1}_i|Y\in[2^{l-j}(2i-2)+1,\,2^{l-j}_{2i}])$
$(l-j)\eps(k_j)$-близко к $M(W,\,d_j)$, \те
$((l-j)\eps(k_j)+\eps(k_j))$-близко к равномерному, откуда с учётом
того, что $\eps(k_j)\leqslant\eps(k_{j-1})$, вытекает утверждение
теоремы.
\end{proof}
\begin{proof}[Доказательство леммы~\ref{concat}]
Достаточно рассмотреть случай, когда $\Prob{Y\in S_1\cup S_2}=1$,
\тк при ограничении на это множество условные распределения не
изменятся. А в этом случае функция $Z=i:Y\in S_i$ будет по условию
($k$,~$\eps$,~0)-селектором для $AB$. По лемме~\ref{srs} $AB$ будет
$\eps$-близко к некоторому распределению $W$ с $\minen(W)\geqslant
k$. В общем случае то же самое будет выполнено для $(AB|Y\in S_1\cup
S_2)$.
\end{proof}

\subsection{Построение итогового экстрактора}
\begin{lemma}[\cite{srinivasan-zuckerman}]\label{sz}
Для некоторой константы $c>1$ и любого $k=\Omega(\log n)$ существует
полиномиально вычислимый ($2k$,~$2^{-k/5}$)-экстрактор $A_k\colon
(n)\times(k)\to(ck)$. Обозначим эту константу $c_{sz}$.
\end{lemma}
Покажем, как можно построить хороший экстрактор, имея хороший
мёрджер.
\begin{lemma}\label{mergext}
Пусть для всех $k\in[k',\,k'']$ существует полиномиально вычислимый
$\eps$-мёрджер $M_k\colon(k)^n\times(d)\to(\alpha k)$, где
$\alpha$\т константа в интервале $(1/c_{sz},\,1)$. Тогда для тех же
$k$ существует полиномиально вычислимый ($k$,
$\poly(n)\cdot\eps$)-экстрактор
$E\colon(n)\times(O(k'\log(1/\eps))+d\log n)\to(\Omega(k))$
\end{lemma}
\begin{proof}
Положим $b=c_{sz}\cdot\alpha$, $k_i=b^ik'\log(1/\eps)$, а $t$\т
минимальное число, что $\sum_{i=1}^t k_i\leqslant k/2$. Определим
$E$ как $E=E_t\merg{M_{t-1}}E_{t-1}\dots\merg{M_1}E_1$, где
\begin{itemize}
\item
$E_i\colon(n)\times(k_i)\to(c_{sz}k_i)$\т
($2k_i$,~$2^{k_i/5}$)-экстрактор из леммы~\ref{sz}.
\item
$M_i\colon(c_{sz}k_{i+1})^n\times(d)\to(k_{i+2})$\т $\eps$-мёрджер,
существующий по предположению леммы. ($k_{i+2}=bk_{i+1}=\alpha\cdot
c_{sz}k_{i+1}$)
\end{itemize}
По теореме~\ref{manyextcomp}, применённой для $d_i=k_i$ и
$s_i=(c_{sz}-b)k_{i-1}$, $E$ является экстрактором. Проверим, что он
имеет нужные параметры. Выберем параметр безопаcности $s=k/2t$,
тогда по выбору $t$ минимальная энтропия $E$ будет равна
$\sum_{i=1}^tk_i+(t-1)s>k$. Очевидно, $t=O(\log n)$, это даёт нужную
оценку на количество случайных битов. Далее,
$k_{t+1}=b^{t+1}k'\log(1/\eps)=\Omega\left(\frac{b^{t+1}-1}{b-1}k'\log(1/\eps)\right)=\Omega(\sum_{i=1}^tk_i)=\Omega(k)$.
Наконец, ошибка $E$ равна
$\sum_{i=1}^t2^{-k_i/5}+(t-1)\eps+(t-1)s2^{-s/3+3}$, что с учётом
того, что $k=O(\log(1/\eps))$, даёт требуемую оценку. Для полного
доказательства осталось заметить, что полиномиальная вычислимость
$E$ также следует из теоремы~\ref{manyextcomp}.
\end{proof}
Сопоставляя результаты теоремы~\ref{mergexist} и
леммы~\ref{mergext}, получаем, что для всех $k$ существует
полиномиальный ($k$,~$\eps$)-экстрактор
$E\colon(n)\times(d\polylog(n)\log(1/\eps))\to(\Omega(k))$, где
$d$\т количество случайных битов, необходимых для работы мёрджера.
Однако, этот экстрактор \лк извлекает\пк ещё не всю минимальную
энтропию исходного распределения, а лишь какую-то её фиксированную
долю. Можно модифицировать алгоритм, чтобы \лк извлечь
случайность\пк полностью. Для этого нам достаточно построить
($k$,~$\eps$)-экстрактор $E\colon(2k)\times(d)\to(k-O(k/\log n))$.

Покажем, как можно увеличить количество извлекаемых битов, используя
один и тот же экстрактор. Пусть экстрактор $E$ \лк извлекает
случайность\пк из всех распределний с минимальной энтропией не
меньше $k$. Что будет, если подать ему на вход распределение с
минимальной энтропией $K\geqslant k$? Можно действовать по
следующему алгоритму: применять экстрактор много раз к одной и той
же строке, каждый раз с новым набором случайных битов $r_i$, пока
суммарная длина выхода на превысит $K-k$. В таком случае с высокой
вероятностью условное распределение $(X|E(x,\,r_1)\dots E(x,\,r_t))$
будет по-прежнему иметь минимальную энтропию не меньше $k$. Более
точно утверждение можно сформулировать в виде двух лемм, доказанных в~\cite{nisan-ta-shma}.
\begin{lemma}\label{lemma411}
Предположим, что для некоторого $k$ существует полиномиально
вычислимый ($k$,~$\eps$)-экстрактор $E_k\colon(n)\times(d)\to(m)$.
Тогда для всяких $K\geqslant k$, параметра безопасности $s>0$ и
$t\in N$ существует полиномиально вычислимый
($K$,~$t(\eps+2^{-s})$)-экстрактор
$E\colon(n)\times(td)\to(\min\{tm,\,K-k-s\})$.
\end{lemma}
\begin{lemma}\label{lemma412}\label{full-extracting}
Предположим, что для всех $k>k'$ существует полиномиально вычислимый
($k$,~$\eps(n)$)-экстрактор $E_k\colon(n)\times(d(n))\to(k/f(n))$.
Тогда для любого $k$ существует полиномиально вычислимый
($k$,~$f(n)\log n(\eps+2^{-d(n)})$)-экстрактор
$E\colon(n)\times(O(f(n)\log n\cdot d(n)))\to(k-k')$.
\end{lemma}
Сопоставив результаты лемм~\ref{mergext} и~\ref{lemma412}, получаем
следующее
\begin{corollary}\label{cor413}
Пусть для всякого $k\geqslant k'=k'(n)$ существует полиномиально
вычислимый $\eps$-мёрджер $M\colon(k)^n\times(d)\to(\alpha k)$, где
$1/c_{sz}<\alpha<1$, Тогда для всякого $k$ существует полиномиально
вычислимый ($k$,~$\poly(n)\eps$)-экстрактор
$E\colon(n)\times(O(k'\log n\log(1/\eps)+\log^2n\cdot d))\to(k)$.
\end{corollary}
(недостающие $k'$ битов можно скопировать непосредственно из
случайных, от этого параметры не ухудшатся)

Теперь построим необходимые мёрджеры. В~\cite{srinivasan-zuckerman}
построены следующие экстракторы:
\begin{lemma}\label{lemma45}
Пусть $k(n)\geqslant n^{1/2+\gamma}$ для некоторого $\gamma>0$.
Тогда для любого $\eps$ существует полиномиально вычислимый
($k(n)$,~$\eps$)-экстрактор
$E\colon(n)\times(O(\log^2n\log(1/\eps)))\to(k^2(n)/n)$
\end{lemma}
На их базе можно построить нужные мёрджеры. А именно, верна
следующая
\begin{lemma}\label{lemma414}
Пусть $b>1$, а $f(k)=f(k(n))$\т такая функция, что
$f(k)\leqslant\sqrt[3]{k}$ и для любого $k\geqslant k_0(n)$
$f(k)\geqslant b\log n$. Тогда для всех $k \geqslant k_0$ существует
$\eps$-мёрджер $M\colon(k)^n\times(\log n\polylog(k)\cdot
f^2(k)\log(1/\eps))\to(k-k/b)$,
\end{lemma}
\begin{proof}
По лемме~\ref{lemma45} существует полиномиально вычислимый
($k/f(k)$,~$\eps$)-экстрактор
$E\colon(2k)\times(O(\log^2k\log(1/\eps)))\to{k/f^2(k)}$. (Здесь
используется неравенство $f(k)\leqslant\sqrt[3]{k}$)

Далее, по лемме~\ref{lemma411} существует полиномиально вычислимый
($k$,~$\poly(k)\cdot\eps$)\д экстрактор $E\colon(2k)\times(O(f^2(k)\log^2k\log(1/\eps)))\to(k-k/f(k))$.

Наконец, по теореме~\ref{mergexist} существует полиномиально
вычислимый $(\log n\poly(k)\cdot\eps)$\д мёрджер $M\colon(k)^n\times(O(\log
n\polylog(k)\cdot f^2(k)\log(1/\eps)))\to(k-\log n\cdot k/f(k))$.
Поскольку $k/f(k)\leqslant k/b\log n$ для всех $k\leqslant k_0$,
имеем $\log n\cdot k/f(k)\leqslant k/b$, что и требовалось.
\end{proof}

Эта лемма позволяет построить экстракторы, которые полностью \лк
извлекают случайность\пк из распределений с высокой минимальной
энтропией. Вначале заметим, что мы можем получить требуемые
мёрджеры:
\begin{corollary}\label{cor415}
Для всех $k\geqslant2^{\sqrt{\log n}}$ существует
$(\polylog(n)\cdot\eps)$\д мёрджер $$M_k\colon(k)\times(\polylog(n)\log(1/\eps))\to(\Omega(k)).$$
\end{corollary}
\begin{proof}
Возьмём $f(k)=\log^c k$ для некоторого $c>2$. Тогда при любом $b$,
$k\geqslant2^{\sqrt{\log n}}$ и достаточно большом $n$ имеем
$\log^ck\geqslant b\log n$, поэтому можно применить
лемму~\ref{lemma414}.
\end{proof}
Подставив этот результат в следствие~\ref{cor413}, получаем
\begin{corollary}\label{cor416}
Для любого $k$ существует
($k$,~$\poly(n)\cdot\eps$)\д экстрактор $$B_k\colon(n)\times(O(2^{\sqrt{\log
n}}\polylog(n)\log(1/\eps)))\to(k).$$
\end{corollary}
Полученный экстрактор \лк извлекает случайность\пк полностью, но
использует сверхполилогарифмическое число случайных битов. Мы можем
уменьшить это число методом композиции экстракторов. Правда, за счёт
этого вновь вырастет требуемая минимальная энтропия. А именно,
\begin{lemma}\label{lemma417}
Пусть $\eps\geqslant2^{-n^{\gamma}}$ для некоторого $\gamma<1$.
Тогда существует $\beta<1$, такая что для всех $k\geqslant n^\beta$
существует полиномиально вычислимый
\extr{E}{k}{\poly(n)\cdot\eps}{n}{\polylog(n)\log(1/\eps)}{\Omega(k)}
\end{lemma}
\begin{proof}
Положим $\delta=(1-\gamma)/2$ и $\beta=1-\delta/2$. Возьмём
$E=B_k\merg{M}E_{sz}$, где
\begin{itemize}
\item
$E_{sz}$\т
($n^\beta$,~$\eps$)\д экстрактор $E\colon(n)\times(O(\log^2n\log(1/\eps)))\to(n^{2\beta-1})$ из
леммы~\ref{lemma45};
\item
$B_k$\т экстрактор из следствия~\ref{cor416};
\item
$M$\т мёрджер из следствия~\ref{cor415}.
\end{itemize}
Поскольку $n^{2\beta-1}=n^\delta n^\gamma=\Omega(2^{\sqrt{\log
n}}\log(1/\eps))$, экстрактор $E$ корректно определён. По
теореме~\ref{twoextcomp} $E$ является полиномиально вычислимым
($k+n^\beta+n^\gamma$,~$\poly(n)\cdot\eps$)-экстрактором
$E\colon(n)\times(\polylog(n)\log(1/\eps))\to(\Omega(k))$. В
частности, если $\minen(X)=\Omega(n^\beta)$, мы извлечём
$\Omega(\minen(X))$ битов, что и требовалось в теореме.
\end{proof}

Наконец, мы можем вновь, сохранив полилогарифмическое число
случайных битов, перейти к произвольной минимальной энтропии и
получить итоговую теорему.
\begin{theorem}\label{lasttheorem}
Для любых $\gamma<1$, $\eps\geqslant2^{-n^\gamma}$ и $k=k(n)$
существует полиномиально вычислимый
\extr{E}{k}{\eps}{n}{\polylog(n)\log(1/\eps)}{k}
\end{theorem}
\begin{proof}
По лемме~\ref{lemma412}, лемма~\ref{lemma417} влечёт существование
полиномиально вычислимого ($n$,~$\poly(n)\cdot\eps$)-экстрактора
$E\colon(2n)\times(\polylog(n)\log(1/\eps))\to(n-n^\beta)$.

Далее, существует константа $c$, зависящая только от $\gamma$, такая
что для всех $\log^c n \leqslant k \leqslant n$ выполнено $\log
n\cdot k^\beta \leqslant k/\bar c$, где $\bar c$ таково, что
$1/c_{sz}<1-1/\bar c<1$. Отсюда по теореме~\ref{mergexist} для
всякого $k$ существует полиномиально вычислимый
\merger{M}{\poly(n)\cdot\eps}{k}{\polylog(n)\log(1/\eps)}{k-k/\bar
c}.

Наконец, по следствию~\ref{cor413} получаем для любого $k$
полиномиально вычислимый
\extr{E}{k}{\poly(n)\cdot\eps}{n}{\polylog(n)\log(1/\eps)}{k}.
Полагая $\eps'=\eps/\poly(n)$, получаем в точности утверждение
теоремы.
\end{proof}

\section{Экстрактор Тревисана}\label{extr-trevisan}
В 1999 году Л.\,Тревисан (Luca Trevisan) в статье~\cite{trevisan} описал совершенно новую конструкцию экстракторов на основе понятий
псевдослучайного генератора, комбинаторного дизайна и кодов, исправляющих ошибки. Эта конструкция гораздо проще всех предыдущих и (в
усиленном виде, описанном в~\cite{rrv}) даёт лучшие параметры экстрактора.

\subsection{Дизайны}
Дизайном называется набор подмножеств некоторого множества, имеющих одинаковый размер и маленькие попарные пересечения. Более строго:
\begin{definition}
Пусть дано множество размера $d$ (напомним, мы обозначали его $[d]$). Семейство множеств $S_1,\dots,S_m\subset[d]$, каждое из которых имеет
размер $l$, называется:
\begin{itemize}
\item
($l$,~$\rho$)\д дизайном (design), если для всех $i\ne j$ выполнено $|S_i\cap S_j|\leqslant\log\rho$;
\item
слабым ($l$,~$\rho$)\д дизайном (weak design), если для всех $j$ выполнено $$\sum\limits_{i<j}2^{|S_i\cap S_j|}\leqslant\rho(m-1);$$
\item
равномерным слабым ($l$,~$\rho$)\д дизайном (uniform weak design), если для всех $j$ выполнено $$\sum\limits_{i<j}2^{|S_i\cap
S_j|}\leqslant\rho(j-1).$$
\end{itemize}
\end{definition}
Очевидно, любой дизайн является равномерным слабым дизайном с теми же параметрами, а равномерный слабый дизайн\т слабым дизайном. Обратное,
разумеется, неверно.

Известны следующие конструкции дизайнов:
\begin{lemma}[\cite{trevisan}, \cite{rrv}]\label{design-exist}
Для всех $l$, $m$ и $\rho=\rho(l,\,m)>1$ существует набор $\mathcal{S}$ множеств $S_1,\dots,S_m\subset[d]$, каждое из которых имеет размер
$l$, который является:
\begin{itemize}
\item
($l$,~$\rho$)\д дизайном при $d=O(l^2m^{1/\rho})/\rho$;
\item
слабым ($l$,~1)\д дизайном при $d=O(l^2\log m)$;
\item
равномерным слабым ($l$,~$\rho$)\д дизайном при $d=O(l^2/\log\rho)$.
\end{itemize}
Более того, этот набор $\mathcal{S}$ можно получить детерминированным алгоритмом за время $O(2^dm)$ для дизайна и за время
$\poly(m,\,d)$ для слабого и равномерного слабого дизайнов. Конструкции для слабого и равномерного слабого дизайнов обладают тем
свойством, что любое их подмножество вида $\{S_1,\dots,S_i\}$ также является соответственно слабым и равномерным слабым дизайном с теми же
параметрами.
\end{lemma}
В работе~\cite{rrv} доказана также оптимальность последних двух оценок.

\subsection{Псевдослучайный генератор Нисана-Вигдерсона}
Следующая конструкция описана в работе~\cite{nw94}. Пусть даны множество $S=\{s_1<\dots<s_l\}\subset[d]$ и $y\in\bcube{d}$. Пусть $y_i$\т
$i$\д ый бит $y$. Обозначим через $y|_S$ слово, образованное битами $y$, лежащими в $S$, \те $y|_S\bydef y_{s_1}\dots y_{s_l}$.
\begin{definition}
Пусть даны функция $f\colon\bcube{l}\to\bcube{}$ и семейство $\mathcal{S}$ подмножеств $S_1,\dots,S_m\subset[d]$ одинакового размера $l$. Тогда
генератором Нисана-Вигдерсона называется функция $NW_{f,\,\mathcal{S}}\colon\bcube{d}\to\bcube{m}$, определённая равенством
$$NW_{f,\,\mathcal{S}}(y)=f(y|_{S_1})\dots f(y|_{S_m}).$$
\end{definition}
В работе~\cite{nw94} показано, что если $f$ является трудновычислимой функцией, а $\mathcal{S}$\т дизайн, то функция $NW_{f,\,\mathcal{S}}$
действительно является псевдослучайным генератором.

\subsection{Коды, исправляющие ошибки}
Обычно под кодами, исправляющими ошибки, понимают отображения из $\bcube{n}$ в $\bcube{\bar n}$, такие что расстояние Хэмминга (\те
количество различающихся битов) между любыми двумя кодовыми (\те лежащими в образе $\bcube{n}$) словами достаточно велико. Нам потребуется
более слабое понятие кодов, допускающих \лк декодирование списком\пк (list-decoding), \те таких, что в окрестности каждого кодового слова
лежит не более полиномиального числа кодовых слов. Мы будем пользоваться кодами, существование которых утверждает следующая
\begin{lemma}\label{ecc}
Для всех $n\in\N$ и $\delta=\delta(n)>0$ существует код $EC\colon\bcube{n}\to\bcube{\bar n}$ для некоторого $\bar n=\poly(n,\,1/\delta)$,
такой что любой (хэммингов) шар в $\bcube{\bar n}$ относительного радиуса $1/2-\delta$ содержит не более $1/\delta^2$ кодовых слов. Более
того, зная центр шара $x$, можно алгоритмически получить список прообразов этих слов (\те декодировать $x$ списком) за время
$\poly(n,1/\delta)$. Можно также считать, что $\bar n$ является степенью двойки.
\end{lemma}
Классический пример такого кода\т конкатенация кодов Адамара и Рида-Соломона, дающая $\bar n=O(n^2/\delta^4)$.

\subsection{Конструкция Тревисана}
Экстрактор Тревисана устроен следующим образом. Получив на вход слова $x\in\bcube{n}$ и $y\in\bcube{d}$, мы сначала закодируем $x$ кодом
$EC$ из леммы~\ref{ecc}, получив слово длины $2^l$, которое будем понимать как таблицу значений некоторой булевой функции $f$. А затем
применим к $y$ генератор Нисана-Вигдерсона, построенный по $f$ и некоторому (слабому) дизайну $\mathcal{S}$. Более формально,
\begin{definition}
Пусть $n$, $d$ и $m$ суть натуральные числа, а $\delta=\delta(n)>0$. Пусть $EC_\delta\colon\bcube{n}\to\bcube{2^l}$\т код из леммы~\ref{ecc}
для выбранного $\delta$, а $\mathcal{S}$\т семейство подмножеств $S_1,\dots,S_m\subset[d]$ одинакового размера $l$. Для $x\in\bcube{n}$
обозначим через $\hat x$ функцию из $\bcube{l}$ в $\bcube{}$, заданную таблицей значений $EC_\delta(x)$. Тогда функцией Тревисана называется
функция $TR_{\delta,\,\mathcal{S}}\colon\bcube{n}\times\bcube{d}\to\bcube{m}$, заданная равенством
\begin{equation}\label{trevisan-def}
TR_{\delta,\,\mathcal{S}}(x,\,y)=NW_{\hat x,\,\mathcal{S}}(y)=\hat x(y|_{S_1})\dots\hat x(y|_{S_m}).
\end{equation}
\end{definition}
При надлежащем выборе параметров функция Тревисана является экстрактором. А именно, верно следующее 
\begin{claim}[\cite{rrv}]\label{trevisan-rrv}
Пусть $k\leqslant n$, $d$ и $m$ суть натуральные числа, а $\eps=\eps(n)>0$. Тогда функция Тревисана $TR_{\delta,\,\mathcal{S}}$,
построенная для $\delta=\eps/4m$ и слабого ($l$,~$\rho$)\д дизайна $\mathcal{S}$, где $\rho=(k-3\log(m/\eps)-d-3)/m$, является
($k$,~$\eps$)\д экстрактором. Более того, она является экстрактором \textit{в сильном смысле}, \те к выданным битам можно приписать
полученные случайные биты с сохранением свойства экстрактора.
\end{claim}
Утверждение~\ref{trevisan-rrv} позволяет получить следующую теорему:
\begin{theorem}[\cite{rrv}]
Пусть $m\leqslant k\leqslant n$ суть натуральные числа, а $\eps>0$. Тогда существует полиномиально вычислимый ($k$,~$\eps$)\д экстрактор
$Ext\colon\bcube{n}\times\bcube{d}\to\bcube{m}$ для
\begin{itemize}
\item
$d=O\left(\frac{\log^2(n/\eps)}{\log(k/m)}\right)$, или
\item
$d=O(\log^2(n/\eps)\log(1/\gamma))$, где $1+\gamma=k/(m-1)$ и $\gamma<1/2$.
\end{itemize}
\end{theorem}
Таким образом, чтобы \лк извлечь всю случайность\пк, нам достаточно $O(\log^2(n/\eps)\log k)$ дополнительных случайных битов, а чтобы
\лк извлечь некоторую постоянную долю случайности\пк \т $O(\log^2(n/\eps))$ битов. Из свойств слабого дизайна следует, что экстрактор
Тревисана будет префиксным.

\section{Конструкция Рейнгольда-Шалтиэля-Вигдерсона}\label{explicit-last}
В 2001 году О.\,Рейнгольд (Omer Reingold), Р.\,Шалтиэль (Ronen Shaltiel) и А.\,Вигдерсон (Avi Wigderson) в статье~\cite{rsw} путём
комбинирования всех
предложенных ранее методов построили наилучшие из известных на сегодняшний день экстракторы. А 
именно, для всех $k$ построены экстракторы с $m=ck$ и $d=O(\log n(\log\log n)^2)$ для произвольной константы $c<1$.
Лемма~\ref{full-extracting} влечёт существование для всех $k$ экстракторов с $m=k$ и $d=O(\log^2 n(\log\log n)^2)$. При построении этих
экстракторов использовался такой инструмент, как конденсеры.
\begin{definition}
Функция $Con\colon\bcube{n}\times\bcube{d}\to\bcube{n'}$ называется ($k$,~$k'$,~$\eps$)-конденсером, если для любого распределения $X$ на
$\bcube{n}$ с $\minen(X)\geqslant k$ существует такое распределение $Y$ на $\bcube{n'}$, что $\minen(Y)\geqslant k'$ и
$\dist(Y,\,Con(X.\,U_d))<\eps$.
\end{definition}
Таким образом, в отличие от экстрактора, конденсер не \лк извлекает случайность\пк, а лишь \лк конденсирует\пк её на некотором меньшем числе
битов.

В работе~\cite{rsw} показано, как построить полиномиально вычислимые конденсеры, и как из них построить экстракторы путём последовательного
применения. При построении конденсеров используются методы, похожие на анализ различных плохих множеств в теореме~\ref{twoextcomp}.

\section{Теорема Мучника}\label{muchnik}

\subsection{Определение колмогоровской сложности}\label{muchnik-pre}

Дадим формальное определение колмогоровской сложности.
\begin{definition}
Пусть $\varphi$\т некоторый алгоритм, который переводит пары двоичных слов в двоичные слова. Тогда условной колмогоровской сложностью
$KS_\varphi(A|B)$ слова $A$ относительно слова $B$ при способе описания $\varphi$ называется длина кратчайшего слова $P$, такого что
$\varphi(P,\,B)=A$.
\end{definition}
Среди всех способов описания найдётся оптимальный. А именно, выполнено следующее
\begin{claim}
Существует такой способ описания $\psi$, что для любого способа описания $\varphi$ и всех слов $A$ и $B$ выполнено $KS_\psi(A|B)\leqslant
KS_\varphi(A|B)+O(1)$. При этом константа в $O(1)$ зависит от $\varphi$, но не от $A$ и $B$.
\end{claim}
\begin{definition}
Условной колмогоровской сложностью $KS(A|B)$ слова $A$ относительно слова $B$ называется $KS_\psi(A|B)$, где $\psi$\т способ описания из
предыдущего утверждения. Колмогоровской сложностью $KS(A)$ слова $A$ называется $KS(A|\Lambda)$, где $\Lambda$\т пустое слово.
\end{definition}
Изучают также колмогоровскую сложность с ограничением на ресурсы.
\begin{definition}
Пусть $\varphi$\т некоторый алгоритм (для многоленточной машины Тьюринга), который переводит пары двоичных слов в двоичные слова. Тогда
условной колмогоровской сложностью
$C^{t,\,s}_\varphi(A|B)$ слова $A$ относительно слова $B$ за время $t$ на зоне $s$ при способе описания $\varphi$ называется длина
кратчайшего
слова $P$, такого что $\varphi(P,\,B)=A$, и $\varphi(P,\,B)$ вычисляется за время $t$ на зоне $s$.
\end{definition}
Утверждение о существовании оптимального способа описания принимает следующий вид:
\begin{claim}
Существует такой способ описания $\psi$, что для любого способа описания $\varphi$ существует такая константа $c$, что всех слов $A$ и $B$
выполнено $C^{ct\log t,\,cs}_\psi(A|B)\leqslant C^{t,\,s}_\varphi(A|B)+c$.
\end{claim}
Любой способ описания, удовлетворяющий этому утверждению, мы будем называть оптимальным. В тех случаях, когда выбор конкретного оптимального
способа описания важен, мы будем его уточнять. Обычно рассматривают сложность либо с ограничением только на время, либо с ограничением
только на память. Условимся, что один индекс означает ограничение на время, а ограничение только на память будем записывать как $C^{\infty,
p}$. 

Для ограничения на ресурсы также рассматривают сложность различения, которая не даёт ничего нового для обычной сложности.
\begin{definition}
Пусть $\varphi$\т некоторый алгоритм, который переводит тройки двоичных слов в $\bcube{}$. Тогда
условной колмогоровской сложностью различения
$CD^{t,\,s}_\varphi(A|B)$ слова $A$ относительно слова $B$ за время $t$ на зоне $s$ при способе описания $\varphi$ называется длина
кратчайшего слова $P$, такого что $\varphi(P,\,U,\,B)=1$, если $U=A$, $\varphi(P,\,U,\,B)=0$ для $U\ne A$, и $\varphi$ вычисляется за время
$t$ на зоне $s$.
\end{definition}
Будем говорить, что $P$\т это программа, принимающая $U$, если $\varphi(P,\,U,\,B)=1$, и отвергающая в противном случае. Безусловная и не
зависящая от способа описания сложности различения вводятся аналогично предыдущим. Также можно определить сложность различения,
релятивизованную относительно некоторого оракула. В работе~\cite{fortnow} доказана следующая оценка:
\begin{theorem}\label{sipser}
Найдётся такой полином $p(n)$, что для любого множества $S$ и всех $A\in S^{=n}=S\cap\bcube{n}$ выполнено
$$
CD^{p,\,S}(A)\leqslant2\log|S^{=n}|+O(\log n).
$$
Более того, найдётся программа, удовлетворяющая этой оценке, спрашивающая оракул только относительно своего входа и отвергающая его, получив
отрицательный ответ.
\end{theorem}

Сложность с ограничением на ресурсы и сложность различения можно рассматривать для вычислительных моделей, отличных от детерминированных
машин Тьюринга. Дадим соответствующие определения, будем считать, что $\varphi$\т оптимальный способ описания.
\begin{definition}
Вероятностной сложностью $CBP^{t,\,s}(A|B)$ называется длина кратчайшей программы $P$, такой что $\Prob{\varphi(P,\,B,\,r)=A}>2/3$, и
$\varphi(P,\,B,\,r)$ работает время $t$ на зоне $s$ для всех $r$.
\end{definition}
Пусть $\varphi_n$\т оптимальный способ описания среди недетерминированных машин Тьюринга.
\begin{definition}
Недетерминированной сложностью $CN^{t,\,s}(A|B)$ называется длина кратчайшей программы $P$, такой что $\varphi_n(P,\,B)$ заканчивает работу
хотя бы для одной ветви алгоритма, возвращает $x$, если заканчивает работу, и работает время $t$ на зоне $s$.
\end{definition}
\begin{definition}
Сложностью Артура-Мерлина $CAM^{t,\,s}(A|B)$ называется длина кратчайшей программы $P$, такой что вероятность того, что
$\varphi_n(P,\,B,\,r)$ заканчивает работу хотя бы для одной ветви и на всех таких ветвях возвращает $x$, больше 2/3, и
$\varphi_n(P,\,B,\,r)$ работает за время $t$ на зоне $s$ для всех $r$.
\end{definition}
Алгоритмы, использующие недетерминизм и случайность, удобно рассматривать как игру двух игроков: Артура и Мерлина. Артур обладает
полиномиальными вычислительными способностями, а Мерлин обладает способностями магическими и может угадывать сертификаты для языков из NP.
При этом и Артур, и Мерлин могут использовать случайность из общего источника. Мерлин стремится вынудить Артура вычислить неверно. Если с
большой вероятностью это у него не получится, то язык лежит в AM.


\subsection{Доказательство теоремы Мучника при помощи экстракторов}

Для доказательства теоремы~\ref{muchnik-th} при помощи экстракторов нам потребуется следующая лемма, доказанная в статье
Л.\,Фортноу~\cite{fortnow}. Будем рассматривать экстрактор как двудольный граф.
\begin{lemma}\label{fortnow}
Пусть существует экстрактор с параметрами $n$, $k$, $d$, $m$, $\epsilon$. Пусть $S$\т подмножество его левой доли размера $K=2^k$.
Назовём \лк плохими\пк элементы правой доли, имеющие больше $2DK/M$ соседей из $S$, и элементы $S$, все соседи которых плохие. Тогда плохих
элементов $S$ не больше чем $2\eps K$.
\end{lemma}
\begin{proof}
Вначале оценим размер множества $Y$ плохих элементов правой доли. По определению экстрактора имеем $\eps>\frac{|E(S,\,Y)|}{D
K}-\frac{|Y|}M$, где $E(S,\,Y)$\т количество рёбер, ведущих из $S$ в $Y$. По выбору $Y$ имеем
$E(S,\,Y)\geqslant|Y|\cdot\frac{2DK}M$, откуда подстановкой получаем $|Y|<\eps M$.
Далее, пусть все соседи множества $X$ попали в множество $Y$. Рассмотрим равномерное распределение на множестве $S$ и применим к нему наш
экстрактор. По свойству экстрактора мы должны получить распределение, $\eps$-близкое к равномерному. Индуцированная экстрактором мера
множества $Y$ не меньше $\frac{|X|}{|S|}$, поэтому 
\begin{equation}\label{lfort}
\frac{|X|}K-\frac{|Y|}M<\eps,
\end{equation}
откуда с учётом $|Y|<\eps M$ получаем $|X|<2\eps K$, что и
требовалось.
\end{proof}
Докажем теорему Мучника в следующей формулировке:

\begin{theorem}\label{muchnik-extractors}
Пусть $A$ и $B$\т произвольные слова. Пусть существует экстрактор с параметрами $n=l(A)$ (\те $n$ равно длине $A$), $k=KS(A|B)$,
$d=\Omega(\log n)$, $m=KS(A|B)$, $\epsilon=1/n^3$. Тогда найдётся слово $X$ длины не более $KS(A|B)+O(1)$, для которого
$KS(X|A)\leqslant d+2\log n+O(1)$ и $KS(A|B,\,X)\leqslant d+2\log n+O(1)$.
\end{theorem}
Таким образом, использование оптимальных экстракторов позволяет получить теорему Мучника в исходной формулировке, а использование одной из
известных конструкций\т теорему, в которой поправки $O(\log n)$ заменены на $\polylog(n)$, но кодирование осуществляется полиномиальным
алгоритмом.
\begin{proof}[Доказательство теоремы~\ref{muchnik-extractors}]

Пусть $E$\т экстрактор, существование которого утверждается в условии теоремы. Будем считать, что выбор $E$ полностью определяется параметрами
$n$ и $m$: либо экстрактор строится по ним полиномиальным алгоритмом, либо появляется первым в каком-нибудь естественном порядке перебора. Будем
рассматривать левую долю $E$ как множество всех слов
длины~$n$ (среди которых есть $A$), а правую\т как множество всех слов длины~$m$, среди которых мы будем искать $X$. Рассмотрим в левой доле
множество $S_B$ всех слов $P$, для которых $KS(P|B)\leqslant m$. Оно имеет размер $O(2^m)$, поэтому по лемме~\ref{fortnow} доля \лк плохих\пк
$P$ (\те все соседи которых имеют больше $2D$ соседей из $S_B$) в нём не превышает $c\eps=c/n^3$ для некоторой константы $c$. Покажем, что $A$
не может быть \лк плохим\пк. Действительно, свойство быть \лк плохим\пк перечислимо: мы можем перечислять множество $S_B$ и
непосредственной проверкой устанавливать, что слово плохое для уже перечисленной части $S_B$. Таким образом, если бы $A$ было плохим, то при
известном $B$ его можно было бы задать числами $m$, $n$ и номером в переборе всех плохих слов, что дало бы $KS(A|B)\leqslant 2\log
n+(m-3\log n+O(1))<m$, однако $KS(A|B)=m$, откуда имеем противоречие. 

Значит, у $A$ есть хороший сосед справа. Обозначим его через $X$ и покажем, что он удовлетворяет требованиям. Действительно,
$KS(X|A)\leqslant2\log n+d+O(1)$: $2\log n$ битов нужно для задания $n$ и $m$, $d$\т для задания номера $X$ среди соседей $A$. Аналогично
при известных $B$ и $X$ для задания $A$ достаточно указать $n$, $m$ и номер $A$ среди соседей $X$ внутри $S$: мы можем перечислять множество
$S$, а значит, и множество соседей $X$ из $S$. Существование требуемого в теореме $X$ установлено, тем самым теорема доказана.
\end{proof}

\subsection{Теорема Мучника в случае нескольких условий}
Исходная теорема Мучника обобщается следующим образом:
\begin{theorem}
Пусть $A$, $B$ и $C$\т произвольные слова сложности не более $n$, а $k\geqslant l$\т натуральные числа, такие что $KS(A|B)\leqslant k$ и
$KS(A|C)\leqslant l$. Тогда найдётся такое слово $X$ длины не более $k+O(\log n)$, что $KS(X|A)\leqslant O(\log n)$, $KS(A|B,\,X)\leqslant
O(\log n)$ и $KS(A|C,\,X|_l)\leqslant O(\log n)$ для префикса $X|_l$ слова $X$ длины $l$.
\end{theorem}
Использование экстракторов позволяет доказать следующую теорему:
\begin{theorem}
Пусть $A$, $B$ и $C$\т произвольные слова сложности не более $n$, а $k\geqslant l$\т натуральные числа, такие что $KS(A|B)\leqslant k$ и
$KS(A|C)\leqslant l$. Пусть существует префиксный экстрактор с параметрами $n$, $k$,
$d$, $m=k$, $\eps=1/n^3$. Тогда найдётся такое слово $X$ длины не более $k+O(\log n)$, что $KS(X|A)\leqslant d+O(\log n)$,
$KS(A|B,\,X)\leqslant d+O(\log n)$ и $KS(A|C,\,X|_l)\leqslant d+O(\log n)$.
\end{theorem}
\begin{proof}
Вновь будем следовать доказательству исходной теоремы, переводя его на язык экстракторов. Как и прежде, можно считать, что длина слова $A$
равна его сложности, то есть $n$.

Вновь будем рассматривать левую долю экстрактора $E$, существование которого утверждается в условии, как множество всех слов длины
$n$, а правую\т как множество всех слов длины $m$, среди которых мы будем искать $X$.
Нам потребуется усиленный вариант леммы~\ref{fortnow}:
\begin{lemma}\label{impr-fortnow}
Пусть существует экстрактор с параметрами $n$, $k$, $d$, $m$, $\epsilon$. Пусть $S$\т подмножество его левой доли размера $K=2^k$.
Назовём \лк плохими\пк элементы правой доли, имеющие больше $2DK/M$ соседей из $S$, и элементы $S$, по крайней мере половина соседей которых
плохие. Тогда плохих элементов $S$ не больше чем $4\eps K$. 
\end{lemma}
\begin{proof}
Доказательство повторяет доказательство леммы~\ref{fortnow} вплоть до неравенства~\eqref{lfort}, которое нужно заменить на
$\frac{|X|}{2K}-\frac{|Y|}M<\eps$, что с учётом $|Y|<\eps M$ повлечёт требуемую оценку $|X|<4\eps K$.
\end{proof}
Вновь рассмотрим множества $S_B$ и $S_C$ слов $P$, имеющих условную сложность не больше $k$ и $l$ относительно $B$ и $C$ соответственно. По
лемме~\ref{impr-fortnow} доля плохих слов в каждом из них не превышает $c/n^3$, поэтому $A$ не может быть плохим ни для одного из
них. Значит, для каждого из слов $B$ и $C$ у $A$ больше половины хороших соседей. Следовательно, хотя бы один сосед будет хорошим для обоих.
Его мы и возьмём в качестве $X$. Свойства $KS(X|A)\leqslant d+O(\log n)$ и $KS(A|B,\,X)\leqslant d+O(\log n)$ доказываются в точности как
раньше, последнее свойство следует из определения префиксного экстрактора.
\end{proof}
Эту теорему можно обобщить на произвольное количество условий:
\begin{theorem}
Пусть $A$, $B_1$, $\dots$, $B_p$\т произвольные слова сложности не более $n$, а $k_1\geqslant\dots\geqslant k_p$\т натуральные числа, такие
что $KS(A|B_i)\leqslant k_i$ для $i=1,\dots,p$. Пусть существует префиксный экстрактор с параметрами $n$, $k$,
$d$, $m=k$, $\eps=1/pn^3$. Тогда найдётся такое слово $X$ длины не более $k_1+O(\log n)$, что $KS(X|A)\leqslant d+O(\log n)$,
$KS(A|B_i,\,X|_{k_i})\leqslant d+O(\log n)$ для $i=1,\dots,p$.
\end{theorem}
\begin{remark}
Поскольку для оптимального экстрактора $d=O(\log (n/\eps))$, то мы получим точность $O(\log n)$ вместо $d+O(\log n)$ лишь для
полиномиальных $p$, как и в исходной теореме из~\cite{muchnik}.
\end{remark}

\subsection{Теорема Мучника для сложности с ограничением на память}\label{muchnik-polyzone-sect}
Немного изменённая конструкция из теоремы~\ref{muchnik-extractors} позволяет распространить утверждение теоремы Мучника на колмогоровскую
сложность с полиномиальным ограничением на память. 

\begin{theorem}\label{muchnik-polyzone}
Пусть $A$ и $B$\т произвольные слова длины не более $n$, а $p$\т произвольное число. Пусть для всех $k\leqslant C^{\infty,\,p}(A|B)$
существует экстрактор $E_k$, вычислимый на зоне $\poly(n)$, с параметрами $n$, $k$, $d$, $m=k$, $\eps=1/n^3$. Тогда найдётся слово $X$ длины не
более $C^{\infty,\,p}(A|B)$, такое что $C^{\infty,\,\poly(n)}(X|A)\leqslant d+O(\log n)$ и $C^{\infty,\,2p+\poly(\log p,\,n)}(A|B,\,X)\leqslant
d+O(\log n)$.
\end{theorem}
Попытка напрямую распространить доказательство теоремы Мучника на сложность с ограничением на память наталкивается на трудность: для
нахождения плохих слов для множества $S_B=\{P\mid C^{\infty,\,p}(P|B)\leqslant C^{\infty,\,p}(A|B)\}$ требуется зона, большая $p$. А тогда
мы не сможем доказать, что $A$ хорошее: утверждения $C^{\infty,\,p}(A|B)\}=k$ и $C^{\infty,\,p'}(A|B)\}<k$ непротиворечивы при $p'>p$.
Однако, эту трудность можно обойти: для хороших $A$ построим хеш-значение старым способом, а для плохих рассмотрим новый экстрактор с меньшей
правой долей, и в нём сделаем (рекурсивно) то же самое. Таким образом получится итеративный процесс, на последнем шаге которого все слова
будут хорошими. Опишем эту конструкцию формально.
\begin{definition}
Будем говорить, что множество $S$ перечислимо на зоне $p$, если существует алгоритм, который по любому числу $z\leqslant|S|$ находит $z$-ый
элемент $S$, при таком входе заканчивает работу на зоне $p$, а при ином входе работает на большей зоне или останавливается без возвращения
результата. Очевидно, это
эквивалентно следующему: некоторый алгоритм, работая на зоне $p$, печатает все слова из $S$ на выходной ленте и останавливается, или выходит
за пределы зоны $p$, при этом занятая на выходной ленте зона не считается.
\end{definition}
\begin{lemma}
Пусть множество $S$, содержащееся в левой доле экстрактора $E_k$, перечислимо на зоне $p$. Тогда
множество $Bad(S)$ плохих для $S$ слов перечислимо на зоне $p+2\log p+\poly(n)$.
\end{lemma}
\begin{proof}
Вначале покажем, как можно перечислить $Bad(S)$ на зоне $p+\log p+\poly(n)$, если $p$ известно. Запуская алгоритм, перечисляющий $S$, и
ограничивая зону его работы числом $p$, мы будем получать слова $P\in S$. Получив очередное слово, проверим, является ли оно плохим, и если
является, то напечатаем на выходной ленте. Покажем, как осуществить такую проверку. Будем последовательно вычислять всех соседей слова $P$ и
проверять, являются ли они плохими. Если все являются, значит $P$ плохое, иначе хорошее. Покажем, как проверить, является ли слово $X$ из правой
части плохим. Снова будем запускать алгоритм, перечисляющий $S$, ограничивая зону его работы числом $p$. У каждого вновь полученного слова
будем вычислять всех соседей и считать, сколько из них совпадают с $X$. Если общее число совпадений превысило $2D$, значит $X$ плохое.
Иначе, \те если алгоритм, перечисляющий $S$, остановился или попытался выйти за пределы зоны, а число совпадений не превысило $2D$, $X$
хорошее. 

Посчитаем использованную зону. Заметим, что в процессе вычисления, занимающего зону $p$, нам не нужно обращаться к другому вычислению,
занимающему такую же зону. Значит, все такие вычисления мы можем проводить на одной и той же зоне $p+\log p$, где $\log p$ нужно для
контроля невыхода за зону $p$. Для вычисления соседей слова из левой доли достаточно зоны $\poly(n)$, для остальных вычислений (переборов и
хранения промежуточных результатов) достаточно зоны $O(n)$. В общей сложности получаем зону $p+\log p+\poly(n)$.

Теперь покажем, как сделать то же самое, не зная заранее $p$, на зоне $p+2\log p+\poly(n)$. Обозначим через $S_q$ множество, которое
перечисляет алгоритм, перечисляющий $S$, с ограничением $q$ на зону его работы. Очевидно, для $q<q'$ верно $S_q\subset S_{q'}$ и
$Bad(S_q)\subset Bad(S_{q'})$. Воспользуемся этим фактом: будем последовательно запускать предыдущий алгоритм для $q=1,\,2,\,3,\dots$.
Если на этапе с ограничением $q$ получено, что слово $P$ плохое, проверим, являлось ли оно плохим на предыдущем этапе (с ограничением
$q-1$), и если не являлось, то напечатаем его на выходной ленте. Таким образом, плохие для $S$ слова будут печататься в другом порядке,
но каждое по одному разу. Рано или поздно алгоритм дойдёт до $q=p$, и все плохие для
$S$ слова к этому времени будут перечислены. По сравнению с предыдущим алгоритмом, потребуется дополнительная зона $\log p$ для организации
перебора всех $q$, и $O(n)$ для хранения промежуточных результатов во время проверки, являлось ли полученное слово плохим на предыдущем
этапе. Таким образом, общая зона составит $p+2\log p+\poly(n)$, что и требовалось.
\end{proof}
\begin{lemma}
Иножество $S_B=\{P\mid C^{\infty,\,p}(P|B)\leqslant k\}$, лежащее в левой доле экстрактора $E_k$, перечислимо на зоне $2p+2\log p+O(n)$.
\end{lemma}
\begin{proof}
Как и в предыдущей лемме, предположим вначале, что перечисляющему алгоритму известно $p$. Будем перебирать все
описания длины $k$ и запускать на них оптимальный способ описания $\psi$ с ограничением на зону $p$. Одновременно будем считать количество шагов
алгоритма $\psi$ и принудительно останавливать его работу, если это количество превысит $2^p$ (это значит, что $\psi$ зациклился). Если
$\psi$ заканчивает работу, то результат его работы лежит в $S_B$, и мы можем включить его в перечисление. Описанный алгоритм требует зоны
$p+\log p$ для моделирования работы $\psi$, зоны $p$ для контроля зацикливания и зоны $O(n)$ для перебора и хранения промежуточных
результатов, всего $2p+\log p+O(n)$.

При неизвестном заранее $p$ применим тот же приём, что и в предыдущей лемме: будем запускать описанный алгоритм для ограничений на зону
$q=1,\,2,\,3\,\dots$ и печатать слова из $S_B$, как только они получены впервые.
\end{proof}
\begin{corollary}
Положим $S_B^0=S_B$ и $S_B^i=Bad(S_B^{i-1})$ при $i\leqslant1$. Тогда при всех $i$ множество $S_B^i$ перечислимо на зоне $2p+\poly(\log p,\,n)$.
\end{corollary}
\begin{proof}
Поскольку на каждом этапе $S_B^i$ уменьшается хотя бы в $n^3/2$ раз, то при $i>k/2\log n$ все $S_B^i$ пусты и, следовательно, перечислимы. Для
меньших $i$ утверждение выполнено по индукции в силу предыдущих двух лемм.
\end{proof}
Теперь докажем саму теорему.
\begin{proof}[Доказательство теоремы~\ref{muchnik-polyzone}]
Пусть $A$ хорошее, \те $A\in S_B\setminus Bad(S_B)$. Тогда у $A$ есть хороший сосед справа, который мы и возьмём в качестве $X$. Если $A$
плохое, то построим новый экстрактор $E_k$ для $k=C^{\infty,\,p}(A|B)-2\log n$. Если $A$ хорошее в новом экстракторе для множества
$S_B^1=Bad(S_B)$, то у него есть хороший сосед, возьмём его в качестве $X$, иначе применим ту же процедуру ещё раз, и так далее, пока не
придём к случаю $k<2\log n$, когда плохих не останется.

Проверим, что условия теоремы выполнены. Для получения $X$ при известном $A$ достаточно знать номер этапа, на котором $A$ будет хорошим, $n$,
$k$, и номер $X$ среди соседей $A$ в экстракторе на этом этапе, всего $d+O(\log n)$ битов. Поскольку экстрактор вычислим на полиномиальной
зоне, $X$ также получается на полиномиальной зоне. Для получения $A$ при известных $B$ и $X$ достаточно знать номер этапа $i$, $n$,
$k$, и номер $A$ среди соседей $X$ в множестве $S_B^i$, всего $d+O(\log n)$ битов. Поскольку $S_B^i$ перечислимо на зоне $2p+\poly(\log
p,\,n)$, то и множество лежащих в нём соседей $X$ перечислимо, причём на такой же зоне: увеличение составит $\poly(n)$ за счёт необходимости
вычислять экстрактор и хранить промежуточные результаты. Таким образом, $C^{\infty,\,\poly(n)}(X|A)\leqslant d+O(\log n)$ и
$C^{\infty,\,2p+\poly(\log p,\,n)}(A|B,\,X)\leqslant d+O(\log n)$, что и требовалось.
\end{proof}

\subsection{Теорема Мучника для сложности с ограничением на время}\label{muchnik-polytime-sect}
Попытка распространить доказательство теоремы Мучника на сложность с полиномиальным ограничением на время наталкивается на серьёзные
трудности: совершенно непонятно, как можно находить плохие слова за полиномиальное время. Однако, другая техника, описанная в
статье~\cite{blvm}, позволяет доказать такую теорему:
\begin{theorem}\label{muchnik-polytime}
Для всякого полинома $p$ найдётся полином $q$, такой что для произвольных слов $A$ и $B$ длины не более $n$, таких что $C^p(A|B)\leqslant
k$, найдётся слово $X$ длины не более $k+O(\log^3 n)$, такое что $C^q(X|A)\leqslant O(\log^3 n)$ и $CAM^q(A|B,\,X)\leqslant O(\log n)$.
\end{theorem}
\begin{proof}
Доказательство опирается не на произвольный экстрактор, а на конструкцию Тревисана и в целом следует доказательству теоремы~3
работы~\cite{blvm}.

По лемме~\ref{design-exist} существует слабый ($l$,~$1$)\д дизайн для $d=O(l^2\log m)=O(\log^3 n)$. Рассмотрим функцию Тревисана
$TR_{\delta,\,1}\colon\bcube{m}\times\bcube{d}\to\bcube{m}$ для такого дизайна, $m=k+d+1$\,\footnote{Фактически это уравнение на $m$, имеющее
решение $m=k+O(\log^3 n)$.} и $\delta=1/8m$. Очевидно, образ множества $S_B=\{u\in\bcube{n}\mid C^p(u|B)\leqslant k\}$ занимает не больше
половины $\bcube{m}$. Обозначим через $\B$ предикат \лк быть в образе $S_B$\пк. Очевидно, для любого $u\in S_B$ имеем
$\Prob{\B(TR_{\delta,\,1}(u,\,U_d))=1}-\Prob{\B(U_m)=1}\geqslant 1/2$, поскольку первая вероятность равна $1$, а вторая не больше $1/2$.
Распишем это более подробно, обозначив через $\hat u$ образ $u$ под действием кода, исправляющего ошибки:
$$
\Prob[y]{\B(\hat u(y|_{S_1})\hat u(y|_{S_2})\dots \hat u(y|_{S_m}))=1}-\Prob[r_1,\dots,r_m]{\B(r_1r_2\dots r_m)=1}\geqslant 1/2,
$$
где $S_1,\dots,S_m$\т используемый дизайн. Очевидно, для некоторого $i$ будет выполнено
\begin{multline*}
\Prob[y,\,r_{i+1},\dots,r_m]{\B(\hat u(y|_{S_1})\dots \hat u(y|_{S_{i-1}})\hat u(y|_{S_i})r_{i+1}\dots r_m)=1}-\\
-\Prob[y,\,b,\,r_{i+1},\dots,r_m]{\B(\hat u(y|_{S_1})\dots \hat u(y|_{S_{i-1}})br_{i+1}\dots r_m)=1}\geqslant 1/2m.
\end{multline*}
Более того, мы можем фиксировать некоторым образом биты $y$ вне $S_i$, чтобы сохранить это соотношение. Обозначим $y|_{S_i}$ через $x$,
тогда все $\hat u(y|_{S_j})$ зависят только от $|S_j\cap S_i|$ общих с $x$ битов. Таким образом, $\hat u(y|_{S_j})$ есть некоторая
функция $\hat u_j(x)$, определяемая $2^{|S_j\cap S_i|}$ битами. Все функции $\hat u_1(x),\dots,\hat u_{i-1}(x)$ можно задать
$\sum_{j<i}2^{|S_j\cap S_i|}$ битами, что по определению слабого дизайна не превосходит $m$. Эти биты,
вычисленные для $u=A$, мы и возьмём в качестве~$X$. Очевидно, при известном $A$ нам достаточно знать $m$, $y$ (вне $S_i$) и $i$,
чтобы их вычислить. Все вычисления будут полиномиальными, поэтому $C^q(X|A)\leqslant O(\log^3 n)$.

Осталось показать, что при известных $B$ и $X$ нам достаточно небольшой дополнительной информации, чтобы полиномиально вычислить $A$
протоколом AM. Как мы уже установили, имеет место неравенство
\begin{multline}\label{hybrid}
\Prob[x\in\bcube{l},\,b,\,r\in\bcube{m-i}]{\B(\hat u_1(x)\dots\hat u_{i-1}(x)\hat u(x)r)=1}-\\
-\Prob[x\in\bcube{l},\,b,\,r\in\bcube{m-i}]{\B(\hat u_1(x)\dots\hat u_{i-1}(x)br)=1}\geqslant 1/2m.
\end{multline}
Обозначим $\hat u_1(x)\dots\hat u_{i-1}(x)br$ через $F(x,\,b,\,r)$. Положим $g_b(x,\,r)$ равным $b$, если $B(F(x,\,b,\,r))=1$, и
равным $1-b$ в противном случае. Покажем, что $\Prob[x,\,b,\,r]{\hat u(x)=g_b(x,\,r)}\geqslant 1/2+1/2m$:
\begin{multline*}
\Prob[x,\,b,\,r]{\hat u(x)=g_b(x,\,r)} = \Prob[x,\,b,\,r]{g_b(x,\,r)=\hat u(x)|b=\hat u(x)}\Prob[x,\,b,\,r]{b=\hat u(x)}+\\
\shoveright{+\Prob[x,\,b,\,r]{g_b(x,\,r)=\hat u(x)|b\ne\hat u(x)}\Prob[x,\,b,\,r]{b\ne\hat u(x)}=}\\
\shoveleft{=\frac12\Prob[x,\,b,\,r]{\B(F(x,\,b,\,r))=1|b=\hat u(x)}+\frac12\Prob[x,\,b,\,r]{\B(F(x,\,b,\,r))=0|b\ne\hat u(x)}=}\\
\shoveleft{=\frac12+\frac12\left(\Prob[x,\,b,\,r]{\B(F(x,\,b,\,r))=1|b=\hat u(x)}-\Prob[x,\,b,\,r]{\B(F(x,\,b,\,r))=1|b\ne\hat u(x)}\right)=}\\
\shoveleft{=\frac12+\frac12\left(\Prob[x,\,r]{\B(F(x,\,\hat u(x),\,r))=1}-\Prob[x,\,r]{\B(F(x,\,1-\hat u(x)))=1}\right)=}\\
=\frac12+\Prob[x,\,b,\,r]{\B(F(x,\,\hat u(x),\,r))=1}-\Prob[x,\,b,\,r]{\B(F(x,\,b,\,r))=1}\geqslant \frac12+\frac1{2m}.
\end{multline*}
Можно положить $b$ равным некоторому $b_1\in\bcube{}$ таким образом, чтобы сохранить это соотношение. Бит $b_1$ можно включить в описание
$u$ при известных $B$ и $X$. Без ограничения общности можно считать, что $b_1=1$, поэтому мы опустим этот индекс в дальнейших рассуждениях.
Покажем, как можно (приблизительно) вычислить $g(x,\,r)$ протоколом Артура-Мерлина.

Если бы мы знали, как можно фиксировать биты $r$, чтобы сохранить соотношение $\Prob[x]{\hat u(x)=g(x,\,r)}\geqslant \frac12+\frac1{2m}$, то
мы могли бы найти $\hat u$ недетерминированным протоколом без всякой случайности. Действительно, мы можем вычислить $g(x,\,r)$
недетерминированным протоколом и получить для всех $x$ строчку $\hat v$, совпадающую с $\hat u$ не меньше чем на доле $\frac12+\frac1{2m}$
от всех битов. Недетерминированность нужна, чтобы вычислять $\B$: в качестве сертификата того, что $\B(Y)=1$, можно предъявить описание
прообраза, лежащего в $S_B$, и номер ребра, ведущего из этого прообраза в $Y$. Поскольку запросы к $\B$ неадаптивные, мы можем в самом
начале указать число $a$ положительных ответов на запросы, тогда нахождение $a$ положительных ответов гарантирует, что все остальные
отрицательные. К сожалению, мы не знаем, как фиксировать $r$, и потому будем выбирать их случайными. Покажем, что в этом случае можно
воспользоваться аналогичным рассуждением.

Скажем, что $r$ даёт $\alpha$\д приближение к $\hat u$, если $\Prob[x]{g(x,\,r)=\hat u(x)}\geqslant\alpha$. Будем отождествлять $g(x,\,r)$
со строкой $z_r$, в которой бит номер $x$ равен $b_1=1$ тогда и только тогда, когда $g(x,\,r)=1$. Количество единиц в строке $z_r$ (\те её
вес, $w(z_r)$) совпадает с количеством слов $x$, для которых $\B(\hat u_1(x)\dots\hat u_{i-1}(x)1r)=1$.

Начнём описывать протокол: вначале Артур выбирает случайные строки $r_1,\dots,r_s$ длины $m-i$ для некоторого полинома $s=s(n)$. Включим в
описание среднее число $\bar a=2^{m-i}\sum_{x,\,r}g(x,\,r)$ положительных значений $\B$ среди всех $r$. Будем требовать, чтобы реальное
число положительных значений $\B$ было близко к его ожиданию $s\bar a$. Покажем, что вероятность этого велика.
\begin{claim}
Для любого $\gamma=\gamma(\bar n,\,m)>0$ найдётся $s=O(\bar n^2/\gamma^2)$, такой что с вероятностью не меньше $3/4$ (по выбору
$r_1,\dots,r_s$) выполнены два условия:
\begin{enumerate}
\item\label{cam-approx}
Доля $1/8m$ от $r_1,\dots,r_s$ даёт $(\frac12+\frac1{4m})$\д приближение к $\hat u$.
\item\label{cam-disperse}
Общее число положительных значений $\B$ на строках $r_1,\dots,r_s$ лежит в пределах $\gamma s$ от ожидания: 
$$\Bigl|\sum\limits_{j=1}^s w(z_j)-s\bar a\Bigr|\leqslant \gamma s.$$
\end{enumerate}
\end{claim}
\begin{proof}
Оценим сверху вероятность того, что одно из этих условий не выполняется. Заметим, что если для некоторого $r$
$$\Prob[x,\,b]{\B(F(x,\,\hat u(x),\,r))=1}-\Prob[x,\,b]{\B(F(x,\,b,\,r))=1}\geqslant \frac1{4m},$$
то $r$ даёт $\left(\frac12+\frac1{4m}\right)$\д приближение к $\hat u$. Назовём $r$ плохим, если оно не даёт $(\frac12+\frac1{4m})$\д
приближения к $\hat u$. Из уравнения~\eqref{hybrid} и неравенства Маркова следует, что
$$\Prob[r]{r\ \mbox{плохое}}\leqslant\frac{1-1/2m}{1-1/4m}<1-\frac1{4m}.$$
Согласно неравенству Чернова, для некоторой константы $c_1>0$ выполнено
$$\Prob[r_1,\dots,r_s]{\mbox{плохих\ больше,\ чем\ }(1-\frac1{8m})s}\leqslant\exp(-c_1s/m^2).$$
Для второго условия, также по неравенству Чернова получаем для некоторой константы~$c_2$
$$\Prob{\Bigl|\frac1s\sum\limits_{j=1}^s w(z_j)-\bar a\Bigr|\geqslant \gamma}\leqslant2\exp(-c_2\gamma^2s/\bar n^2).$$
Взяв $s=c_3\bar n^2/\gamma^2$ для достаточно большой константы $c^3$, получаем верхнюю оценку $1/8$ в каждом случае, откуда следует исходное
утверждение.
\end{proof}
После выбора слов $r_1,\dots,r_s$ Артур просит у Мерлина $s\bar a-s\gamma$ сертификатов принадлежности различных слов к $\B$ и проверяет их.
Если хоть один из них ложный, то вычисления заканчиваются без возвращения ответа. Иначе Артур вычисляет строки $z_{r_1}',\dots,z_{r_s}'$,
где $x$\д ый бит строки $z_{r_j}'$ равен $1$ тогда и только тогда, когда Мерлин предоставил сертификат того, что $\B(F(x,\,1,\,r_j))=1$.
Покажем, что с высокой вероятностью вне зависимости от предоставленных Мерлином сертификатов доля $\frac1{16m}$ строк $z_{r_1}',\dots,z_{r_s}'$
даёт $\left(\frac12+\frac1{8m}\right)$\д приближение к $\hat u$.
\begin{claim}
Если $r_1,\dots,r_s$ удовлетворяют условиям предыдущего утверждения для $\gamma=\bar n/256m^2$, то вне зависимости от предоставленных
Мерлином сертификатов доля $\frac1{16m}$ строк $z_{r_1}',\dots,z_{r_s}'$ даёт $\left(\frac12+\frac1{8m}\right)$\д приближение к $\hat u$.
\end{claim}
\begin{proof}
По предположению, число положительных значений $\B$ для $r_1,\dots,r_s$ лежит между $s\bar a-s\gamma$ и $s\bar a+s\gamma$, при этом $s\bar
a-s\gamma$ из них известны Артуру. Поскольку Артур проверил переданные сертификаты, то в тех позициях, где в $z_{r_j}'$ стоит $1$, в
$z_{r_j}$ тоже стоит $1$. Общее количество различий $z_{r_j}'$ от $z_{r_j}$ не превосходит $2s\gamma$, поэтому число таких $r_j$, где строки
различаются в $t$~битах, не больше $2s\gamma/t$.

По предположению, доля $1/8m$ строк $z_{r_j}$ даёт $\left(\frac12+\frac1{4m}\right)$\д приближение к $\hat u$. Положив $t=\bar n/8m$ и
$\gamma=\bar n/256m^2$, получим, что по крайней мере доля $\frac1{8m}-\frac{2\gamma}t=\frac1{16m}$ строк $z_{r_j}'$ совпадают с $\hat u$ не
меньше чем на доле $\frac12+\frac1{4m}-\frac1{8m}$ битов.
\end{proof}
Соединив эти утверждения для достаточно большого полинома $s$, например, $s=\omega(m^4)$, получаем, что с вероятностью по крайней мере $3/4$
по меньшей мере доля $\frac1{16m}$ строк $r_1,\dots,r_s$ даёт $\left(\frac12+\frac1{8m}\right)$\д приближение к $\hat u$. Всякое конкретное
$r_j$ даёт $\left(\frac12+\frac1{8m}\right)$\д приближение лишь для тех кодовых слов, которые совпадают с $z_{r_j}$ по крайней мере на доле
$\frac12+\frac1{8m}$ позиций. По свойству кода таких слов не больше некоторого полинома $q(m)$.

Назовём $\hat v$ кандидатом ($\hat v\in Cand$), если не меньше доли $1/32m$ от всех $r\in\bcube{m-i}$ дают
$\left(\frac12+\frac1{8m}\right)$\д приближение для $\hat v$. Всего кандидатов не больше, чем $32mq$, поскольку число рёбер, соединяющих
кандидаты и $r_j$, не больше $2^{m-i}q$ и не меньше $|Cand|\cdot 2^{m-i}/32m$. По теореме~\ref{sipser} существует программа $p_1$ длины не
больше $2\log(32mq)=O(\log n)$, принимающая $\hat u$ и отвергающая все остальные кандидаты $\hat v$.

Построим список всех кодовых слов $\hat v$, совпадающих с одним из $z_{r_1}',\dots,z_{r_s}'$ хотя бы на доле $\frac12+\frac1{8m}$ от всех
позиций. Затем удалим все слова, встретившиеся меньше $s/16m$ раз. С вероятностью, превосходящей $2/3$, $\hat u$ осталось в списке, и все
оставшиеся в списке слова являются кандидатами. В таком случае из всех оставшихся элементов списка $p_1$ примет $\hat u$ и только его.
Поскольку список получен за полиномиальное время, нам не нужно обращаться к оракулу.

Таким образом, мы получили $\hat u$, зная $\hat u_1,\dots,\hat u_{i-1}$ и следующую дополнительную информацию: номер индекса $i$, бит $b_1$,
среднее число $\bar a$ положительных значений $\B$ и различающую программу $p_1$. Заметим, что нам достаточно знать приближение к $\bar a$,
заданное $O(\log n)$ битами (так чтобы была определена целая часть $s\bar a$), поэтому вся дополнительная информация занимает $O(\log n)$
битов.
\end{proof}
\begin{remark}
Теорема была сформулирована для длины $X$, не превосходящей $k+O(\log^3 n)$ и сложности $CAM^q(A|B,\,X)\leqslant O(\log n)$. Разумеется,
можно \лк перекинуть\пк лишние биты из $X$ в программу алгоритма AM и получить теорему для $X$ длины $k$ и $CAM^q(A|B,\,X)\leqslant O(\log^3
n)$.
\end{remark}

\subsection{Предмет дальнейших исследований} 
Основным вопросом теории экстрактором остаётся построение полиномиально вычислимого оптимального экстрактора. Его решение позволит (помимо
прочих замечательных приложений) получить эффективный вариант теоремы Мучника с исходной точностью. Более слабой задачей является построение
оптимального экстрактора, вычислимого на полиномиальной зоне. Насколько известно автору, эта задача не решена и даже не исследовалась.
Поскольку в доказательстве теоремы Мучника использовалось не свойство экстрактора, а некоторое его следствие, возможно, есть способ
построить \лк псевдоэкстракторы\пк с оптимальными параметрами, для которых выполнено это следствие, но не основное свойство. Также остаётся
вопрос, верна ли теорема Мучника для полиномиального ограничения на время в формулировках, отличных от теоремы~\ref{muchnik-polytime}.

\section{Благодарности}
Автор благодарен А.\,Е.\,Ромащенко и А.\,Шеню за постановку задачи, плодотворное обсуждение и конструктивную критику, а также
А.\,Ю.\,Румянцву за обсуждение результата для сложности с ограничением на память. Кроме того, автор благодарен
оргкомитету Турнира Городов и лично С.\,А.\,Дориченко за включение в состав варианта осеннего тура 2005 года задачи, составленной по мотивам
утверждения~\ref{equiv}.

\end{document}